\documentclass{article}

\usepackage{amsmath,amssymb,amsfonts,dcolumn,color,graphicx,graphics,latexsym,placeins,epsfig}
\usepackage{graphicx}  
\usepackage{amsmath}   
\usepackage{amssymb}   
\usepackage{bm} 
\usepackage{dcolumn}
\usepackage{color}
\usepackage{footmisc}
\usepackage{mathrsfs}
\usepackage{amsfonts}
\usepackage{varioref}
\usepackage{slashed}
\usepackage{footnote}
\usepackage[belowskip=-17pt,aboveskip=15pt]{caption}
\usepackage{amsmath}
\usepackage{makeidx}
\usepackage{amssymb}
\usepackage{graphicx}
\usepackage{bmpsize}

\usepackage[margin=1.45in]{geometry}

\usepackage{graphicx}
\usepackage{dcolumn}
\usepackage{amsmath,amssymb,mathtools}
\usepackage{epstopdf}
\usepackage{verbatim}
\usepackage[colorlinks=true, citecolor=blue, urlcolor = blue, linkcolor= blue, 
bookmarks=true]{hyperref}
\def\be{\begin{equation}}
\def\ee{\end{equation}}

\labelformat{section}{Section #1} 
\labelformat{subsection}{Section #1} 
\labelformat{subsubsection}{Section #1}
\labelformat{subsubsubsection}{Section #1}
\labelformat{equation}{Eq.~(#1)} 
\labelformat{figure}{Fig.~#1} 
\labelformat{subfigure}{Fig.~\thefigure#1} 
\labelformat{table}{Tab.~#1} 
\labelformat{appendix}{Appendix #1}
\vspace{25pt}
\title{\bf Decoherence and entropy generation in an open quantum scalar-fermion system with Yukawa interaction}
\author{$^{1,2}$Sourav Bhattacharya\footnote{sbhatta.physics@jadavpuruniversity.in},\,\, $^2$Nitin Joshi\footnote{2018phz0014@iitrpr.ac.in}~~and~$^2$Shagun Kaushal\footnote{2018phz0006@iitrpr.ac.in}\\
\small{$^1$Relativity and Cosmology Research Centre, Department of Physics, Jadavpur University, Kolkata 700 032, India\footnote{On lien from IIT Ropar, Punjab, India}}\\
\small{$^2$Department of Physics, Indian Institute of Technology Ropar, Rupnagar, Punjab 140 001, India}\\}

\begin{document}

\maketitle

\begin{abstract}
\noindent
We have studied the decoherence mechanism in a  fermion and scalar quantum field theory with the Yukawa interaction in the Minkowski spacetime, using the non-equilibrium effective field theory formalism appropriate for open systems. The scalar field is treated as the system whereas the fermions as the environment. As the simplest realistic scenario, we assume that an observer measures only the Gaussian 2-point correlator for the scalar field. The cause of decoherence and the subsequent entropy generation is the ignorance of information stored in higher-order correlators, Gaussian and non-Gaussian, of the system and the surrounding.
Using the 2-loop 2-particle irreducible effective action, we construct the renormalised Kadanoff-Baym equation, i.e., the equation of motion satisfied by the 2-point correlators in the Schwinger-Keldysh formalism. These equations  contain the non-local self-energy corrections.  We then compute the statistical propagator in terms of the 2-point functions.  
Using the relationship of the statistical propagator with the phase space area, we next compute the von Neumann entropy, as a measure of the decoherence or  effective loss of information for the system. We have obtained the variation of the entropy with respect to various relevant parameters. We also discuss the qualitative similarities and differences of our results with the scenario when  both the system and the environment are scalar fields. \end{abstract}
\newpage

\tableofcontents

\section{Introduction}
A fundamental mark of quantum mechanics is the superposition principle, which leads naturally to the phenomenon of coherence and interference, as appropriate for isolated or ideal quantum systems. In contrast, realistic quantum systems are never completely isolated from their environment. For such {\it open quantum systems}, there is always a source of decoherence due to the interaction or entanglement of the system with the environment or the surrounding \cite{decoherence1, decoherence2, decoherence3}. Such entanglement may influence what we  observe locally 
upon measurement of  the system, even from a classical point of view. Quantum decoherence leads to quantum to classical transition and ensures consistency between quantum and classical predictions for the observed system~\cite{decoherence4}.

Quantum decoherence is rapidly gaining
interest in the research community, chiefly in the context of interacting quantum field theories~\cite{Calzetta Hu, Calzetta Hu1}. Decoherence should be closely related to the loss or ignorance of information of an open quantum system and as we have mentioned above, is usually formulated in the setup of system plus environment. It can be characterised via various correlations,  like the mutual information, discord and the entanglement entropy. Decoherence can be studied in many different scenarios. A general model of decoherence for a non-relativistic quantum particle interacting with a weak stochastic gravitational perturbation was studied in~\cite{gravity, gravity1}. Decoherence generation from an accelerated time-delay source for an inertial observer is studied in~\cite{acceleration}. The decoherence via the gravitational
interaction of the dark matter with its environment, consisting  of ordinary matter is analysed in \cite{darkmatter}. Perhaps one of the most interesting outcome of the decoherence mechanism could be the classicalisation of the primordial inflationary quantum field theoretic  perturbations, leading to the large scale structures in the sky we observe today.  We refer our reader to~\cite{cosmology, cosmology1, DC, DC1, DC2, DC3, Janssen:2007ht, Friedrich, Markkanen:2016vrp, Hu:1992xp, Hu:1990cr} and references therein for some such discussions.

As of the issue of quantifying decoherence, different strategies exist in the non-equilibrium quantum field theory, e.g. \cite{Calzetta Hu, NEQFT, DTLN, noise, buyanovsky, FCL} and also the references therein. In the conventional approach to compute decoherence, one traces out the inaccessible environmental degrees of freedom to obtain a reduced density matrix of the system, which in general is a mixed one. Using this reduced density matrix one quantifies the decoherence generated in the system in terms of the von Neumann entropy, e.g. \cite{buyanovsky, buyanovsky1,  HSSS, SSSS, Global_3, SHN:2020,  SN:2021, SK:2022, FA, entropy, entropy1, Berges, Schmidt}. Another way to tackle the decoherence problem is to consider the master equation approach \cite{Shaisultanov:1995cf, mastereq, mastereq1, master3}. In this work, we shall instead be interested in using the correlator approach proposed in~\cite{JFKTPMGS, koksma, kok}, in order to compute the decoherence in terms of the von Neumann entropy due to observer's lack of ability to know all correlation functions of the system and the environment, eventually characterising the effective loss of information for the system.

Precisely, there can be 
$n$-point correlators which can be generated using the $n$-particle irreducible effective action~\cite{Calzetta Hu, Eaction, Calzetta:1986cq}, capturing the information of interaction between the system and its surrounding. In a realistic scenario however, it is certainly impossible for an observer to quantify the correlators of all orders exactly.  Thus one needs to consider  a practical scenario allowing us to compute only some finite order correlators such as the $2$-point or $4$-point Gaussian ones. The ignorance of the higher-order Gaussian and non-Gaussian correlators of the system and surrounding leads to the lack of information, generation  of quantum decoherence and hence  a non-zero entropy defined in some suitable manner. 
	However, we also note that due to progresses in cold atom experiments, simulating quantum field theory models and measuring higher-order
	correlations becomes an experimentally relevant problem. For example in \cite{prl}, multipoint correlation functions in both in and out of equilibrium quantum field theories have been experimentally studied to show the deviation of Gaussianity due to the presence of interaction. 

In this work we have  considered an open quantum field theory at zero temperature, consisting of bosons and fermions, in order to study the decoherence and subsequent entropy generation  via the approach proposed in~\cite{JFKTPMGS, koksma, kok}. Specifically, we have considered a scalar field coupled to fermions via  the Yukawa interaction,  treating the scalar as the system and the fermions as the surrounding. We shall work in the flat spacetime, and the present work is to be understood as a warm up exercise before we attempt  this problem in the context of the early inflationary universe paradigm, where the late time non-perturbative secular effects may be present.  Some earlier analysis on open quantum systems with scalars and fermions can be found in~\cite{Nusseler:2019ghw, Enqvist:2004pr, Anirban, Lankinen:2019vgv}.

The rest of the paper is organised as follows. In \ref{section : The model}, we have described the model we are considering and the assumptions we are making along with their justifications.  Further, following \cite{JFKTPMGS, koksma, kok} and references therein, we have defined the von Neumann entropy of the system in terms of the phase space area. In \ref{Propagators in the Schwinger-Keldysh Formalism}, we have outlined the in-in or the Schwinger-Keldysh or the closed time path formalism which will be required to compute the relationship between the statistical propagator with that of  the Wightman functions, i.e. the two point functions relevant for our computations. In \ref{The Kadanoff-Baym equations} we have computed the 2-loop 2-particle irreducible (2PI) effective action using the Schwinger-Keldysh formalism and have found out the Kadanoff-Baym equations for our model. These are basically the equations of motion for the loop corrected two point functions in an  interacting quantum field theory, containing self energy corrections. In \ref{Renormalising the Kadanoff-Baym Equations}, we have renormalised the Kadanoff-Baym equations by renormalising the self-energy of the scalar field. Finally in \ref{Phase space area and Entropy}, we have computed the phase space area, the statistical propagator and the entropy of the system, which can be thought of as a quantifier of the decoherence generated on the system due to the interaction with the surrounding and the subsequent ignorance  of the higher order correlators. The variation of the entropy and phase space area with respect to the rest mass of the system as well as the Yukawa coupling strength is  obtained. Finally, in \ref{Conclusion} we conclude our work. The fermions will be taken to be massless, for the sake of simplicity of computation. We shall use the technical formalism of~\cite{koksma, kok}, developed in the context when both system and environment are scalar fields.

We shall work with the mostly positive signature of the metric in $d=(4-\epsilon)$-dimensional ($\epsilon=0^+$) Minkowski spacetime and will set $c=\hbar=1$ throughout.

\section{The basic setup}
\label{section : The model}

We consider a hermitian massive scalar field $\phi(x)$ coupled to fermions by the Yukawa interaction, 
\begin{equation}\label{action:tree1}
S = \int \mathrm{d}^{\scriptscriptstyle{d}}\!x \left[-\frac{1}{2} (\partial_\mu\phi)
(\partial^{\mu} \phi)  - \frac{1}{2} m^{2}\phi^{2}-\frac{i}{2}{\bar{\psi}}\gamma^{\mu}\partial_{\mu}\psi-g {\bar{\psi}}\psi \phi\right]
\end{equation}
Since we are working with the mostly positive signature of the metric, the $\gamma^{\mu}$'s satisfy the anti-commutation,
$$[\gamma^{\mu},\gamma^{\nu}]_+= -2\eta^{\mu\nu} {\bf I}$$
As we have mentioned earlier, the scalar $\phi(x)$ will play the
 role of the system, interacting with the environment $\psi(x)$, $\overline{\psi}(x)$.  We assume that  the environment is at zero temperature and is  in its
vacuum state.  
 \begin{figure}
     \centering
     \includegraphics[scale=.43]{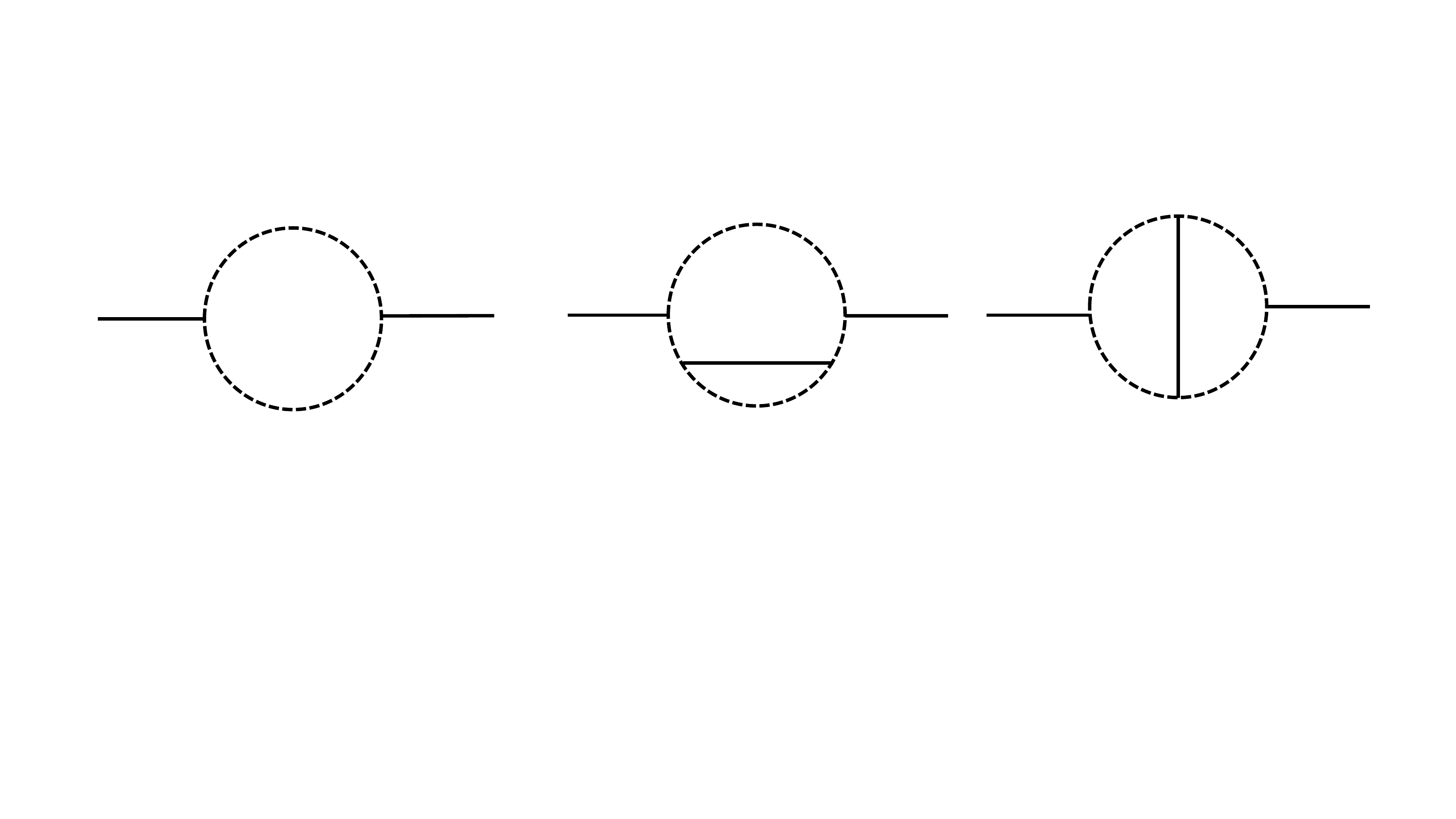}
     \vspace{-40mm}
     \caption{\small \it Self energy diagrams for the scalar field at one and two loop order for the Yukawa coupling.  The broaken lines denote fermions whereas the solid lines denote the scalar. We  shall restrict our computations to  ${\cal O}(g^2)$ only. See main text for discussion. }
     \label{figa}
 \end{figure}

We assume that our observer (observing the system) can measure two point correlation functions only. Note that a tree level two point function is essentially  Gaussian. A quantum  corrected such function may or may not be Gaussian.  Higher  correlations containing the effect of interaction, such as the three point correlators, are  non-Gaussian.  We have to  work on perturbative corrections to the self energy and will restrict ourselves to one loop, ${\cal O}(g^2)$. Note also that the two loop ${\cal O}(g^4)$ diagrams of \ref{figa} contains backreaction of the system on the environment,  which will be ignored. Restricting ourselves only to one loop  seems justified at least when the Yukawa coupling constant $g$ is not too strong and second, in particular when there is no secular effect at late times. Note also that since the fermions are taken to be massless, there can be decay of the system (i.e., the scalar) into fermion-anti-fermion pairs starting at ${\cal O}(g^2)$. However, this decay involves three point correlators between the system and the surrounding, with on shell process also going on within the surrounding which is unobserved by the observer.  Hence we shall ignore such decay in our computations. As we have stated in the preceding Section, the lack of observer's ability to measure higher order correlators due to the system-environment interaction essentially leads to lack of information for the system and it decoheres. This lack of information or decoherence will be quantified  by the von Neumann entropy, as follows. 

 A quantum system can be represented by the density matrix
 that holds all information about the system. One can
 calculate various correlators from the density matrix. In particular, different two point correlators are given as \cite{EKP}
\begin{eqnarray}
\langle\phi(x)\phi(x^\prime)\rangle &=&  \mathrm{Tr}\left[{\rho}(t)
\phi(x)\phi(x^\prime) \right]
\nonumber\\
\langle \pi(x)\pi(x^\prime)\rangle &=&  \mathrm{Tr}\left[\rho(t)
\pi(x)\pi(x^\prime) \right]
\nonumber\\
\frac{1}{2}\langle[\phi(x),\pi(x^\prime)]_+\rangle &=&  \frac{1}{2}\mathrm{Tr}[{\rho}(t)
[\phi(x),\pi(x^\prime)]_+]:=\partial_{t'}F_{\phi}(x,x')
\label{p}
\end{eqnarray}
where the first one is the Wightman function,  $\pi(x)=\dot{\phi}(x)$ is the momentum conjugate to the scalar field and  the quantity  $F_{\phi}(x,x')$ is called the statistical propagator defined as
\begin{equation}\label{statisticalpropagator}
F_{\phi}(x,x') := \frac{1}{2} \mathrm{Tr} \left(
\rho(t)[\phi(x), \phi(x')]_+ \right)=
\frac{1}{2} \mathrm{Tr} \left[ \rho(t) (
\phi(x)\phi(x') + \phi(x')\phi(x)) \right]
\end{equation}
Thus the statistical propagator is basically the Wightman function symmetrised in $x$ and $x'$. For a given density operator $\rho(t)$ (pure or mixed), the statistical propagator tells us about how the states are occupied. One can also relate this to the average particle density.
We shall focus only on the above two point correlators and their quantum corrections, as a practical scenario.

In the spatial momentum space, the statistical propagator reads,
\begin{equation}\label{statpropagatorFourier}
F(|\vec{k}|,t,t')=\int d^3\Delta\vec{x} \, F_{\phi}(t,\vec{x},t',\vec{x'})
e^{-\imath \vec{k}\cdot\Delta \vec{x}}
\end{equation}
One also defines the phase space area for each spatial momentum mode as the Fourier transform of the quantity 
$$4\Big[\langle\phi(x)\phi(x^\prime)\rangle \langle\pi(x)\pi(x^\prime)\rangle-\left(\frac{1}{2}\langle[\phi(x),\pi(x^\prime)]_+\rangle\right)^2 \Big]_{t=t'},$$
given by
\begin{equation}\label{deltaareainphasespace}
\Delta_{|\vec{k}|}^{2}(t)=4 \left[
F(|\vec{k}|,t,t')\partial_{t}\partial_{t'}F(|\vec{k}|,t,t') -
(\partial_{t}F(|\vec{k}|,t,t'))^{2} \right]_{t=t'}
\end{equation} 
where we have used the equal time limits of \ref{p} and \ref{statpropagatorFourier}. In order to see the analogy of this construction with that of ordinary quantum mechanics, let us recall the generalised uncertainty relation (with $\hbar=1$),
\begin{eqnarray} \label{un}
\left\langle q^{2}\right\rangle\left\langle p^{2}\right\rangle-\left[\left\langle\frac{1}{2}[q, p]_+\right\rangle\right]^{2}=\frac{1}{4}\qquad  \text { (pure state) }  \nonumber\\
\left\langle{q}^{2}\right\rangle\left\langle p^{2}\right\rangle-\left[\left\langle\frac{1}{2}[q, p]_+\right\rangle\right]^{2}>\frac{1}{4} \qquad  \text { (mixed state)}
\end{eqnarray}
Combining the two above, we write
\begin{eqnarray} \label{unc}
\left\langle q^{2}\right\rangle\left\langle p^{2}\right\rangle-\left[\left\langle\frac{1}{2}[q,p]_+\right\rangle\right]^{2}=\frac{\Delta^2}{4}
\end{eqnarray}
Thus  the quantity $\Delta \geq 1$, interpreted as the phase space area, can be thought of as a measure of the impurity of the quantum state.  One plausible way to understand the increase in $\Delta$ is the transfer of momentum between the system and the environment, thereby increasing the momentum uncertainty.  Putting these all in together, the loss or ignorance of information due to the inaccessibility of all the correlations in the system, environment and between them, is characterised via the von Neumann entropy     for our field theoretic continuum system~\cite{EKP},
\begin{equation}\label{entropy}
S_{|\vec{k}|}(t) = \frac{ \Delta_{|\vec{k}|}(t)+1}{2}
\ln\left(\frac{\Delta_{|\vec{k}|}(t)+1}{2}\right) - \frac{
\Delta_{|\vec{k}|}(t)-1}{2} \ln\left(\frac{\Delta_{|\vec{k}|}(t)-1}{2}\right) 
\end{equation}
One can also relate the phase space area with the statistical particle number density per mode as
\begin{equation}\label{particlenumber}
n_{|\vec{k}|}(t) = \frac{ \Delta_{|\vec{k}|}(t)-1}{2}
\end{equation}
We shall see below that $\Delta_{|\vec k|}(t)$ becomes identity in the absence of the Yukawa interaction.  

For our purpose, we need to compute the various two-point functions in the in-in or the Schwinger-Keldysh formalism, which we outline below.

\subsection{Propagators in the in-in formalism}
\label{Propagators in the Schwinger-Keldysh Formalism}

\begin{figure}
        \begin{center}
\includegraphics[scale=.37]{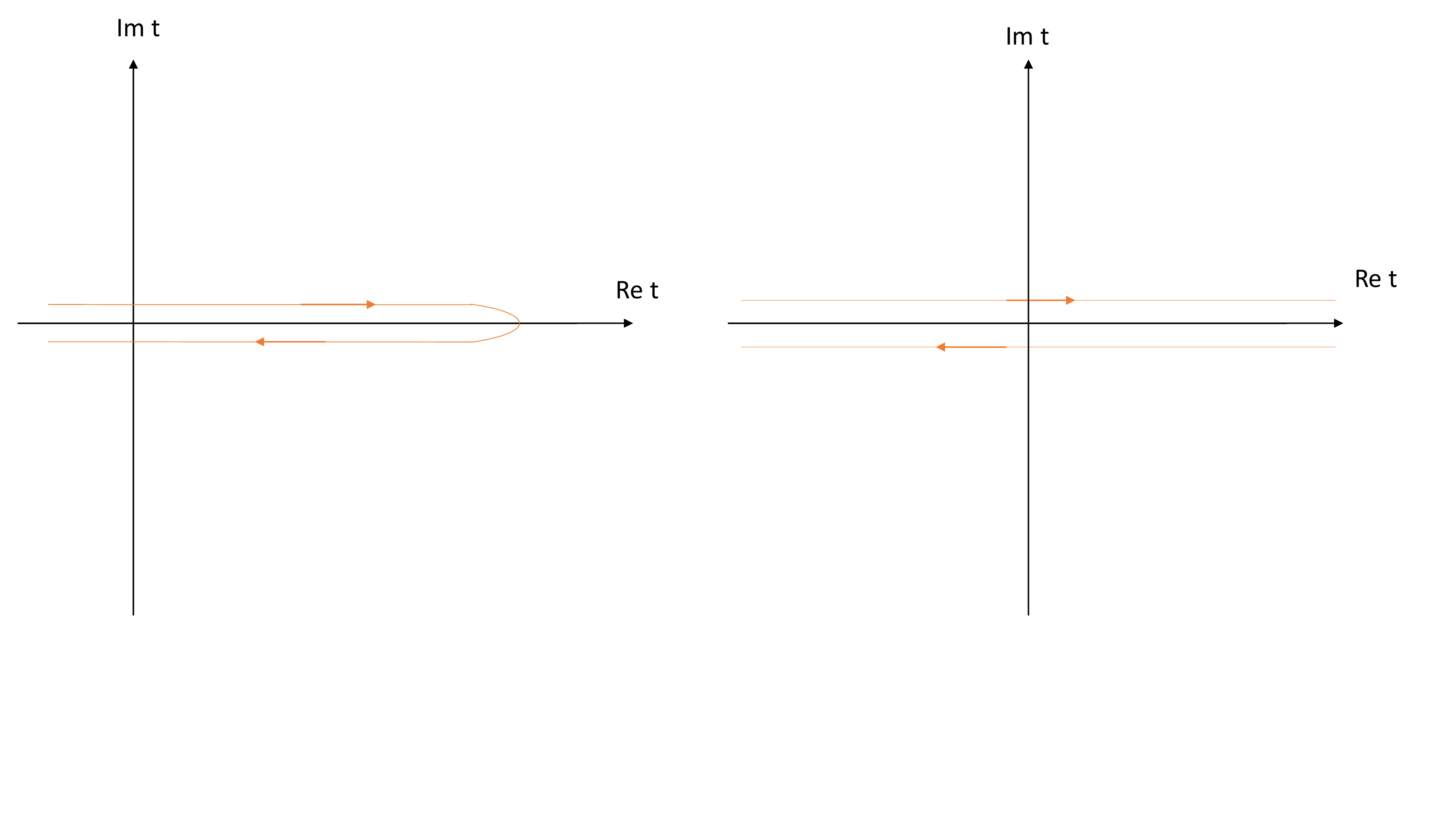}
\vspace*{-20mm}
   {\em \caption{\small \it Schwinger-Keldysh contours with finite and infinite initial and  final times.
   \label{fig:schwingercontour1} }}
        \end{center}
\end{figure}

 The Schwinger-Keldysh or the in-in formalism  is useful for studying the quantum dynamics of a system in a non-equilibrium scenario~\cite{Schwinger:1960qe, Keldysh:1964ud}. Using this, one can meaningfully compute the causal expectation value of an operator with respect to some suitable initial state. Precisely, one follows the time evolution
 of an operator from some initial state at $t = t_0$, without having the knowledge of appropriate late time states. Evolution of an operator from
$t_0$ to $t$ requires a time-ordered evolution and then it is brought back to the initial time following an anti-time ordered evolution $t$ to $t_0$, as depicted in \ref{fig:schwingercontour1}.
 
Thus in the in-in formalism, the expectation value of an operator 
 $O(t)$ with respect to some initial density operator 
$\rho(t_{0})$ (defined in the Heisenberg picture) is given by,
\begin{equation} \label{expectationvalues}
\langle O(t) \rangle =
\mathrm{Tr}\left[\rho(t_{0})O(t)\right] =
\mathrm{Tr}\left[ \rho(t_{0}) \left\{ \overline{T}
\exp\left(\imath \int_{t_0}^t \mathrm{d}t^\prime 
H(t^\prime)\right) \right\} O(t_{0}) \left\{ T \exp
\left(-\imath \int_{t_0}^t \mathrm{d} t'' H(t'')
\right) \right\}\right]
\end{equation}
 where $\overline{T}$ stands for the anti-time  ordering,
and $ H(t)$ is the Hamiltonian. The  generating functional for the corresponding path integral  for \ref{action:tree1}, subject to \ref{fig:schwingercontour1} is given by

\begin{eqnarray}\label{Z:inin}
&& {\cal Z}[J_{+}^{\phi}, J_{-}^{\phi}, J_{+}^{\psi},
J_{-}^{\psi}, \rho(t_0)] \nonumber\\
&& \quad  \!=\! \int \! {\cal D}\phi^{+}_{0}{\cal D}\phi^{-}_{0}
[{\cal D}\psi^{+}_{0}] [{\cal D}\psi^{-}_{0}] \langle\phi^{+}_{0},
\psi^{+}_{0}| { \rho}(t_{0})|\phi^{-}_{0}, \psi^{-}_{0}
\rangle \!\int_{\phi_{0}^{+}}^{\phi_{0}^{-}}\! {\cal
D}\phi^{+}{\cal D}\phi^{-}
\delta[\phi^{+}(t_{f}\!)-\phi^{-}(t_{f}\!)]\!
\int_{\psi_{0}^{+}}^{\psi_{0}^{-}} \![{\cal D}\psi^{+}][{\cal
D}\psi^{-}] [\delta[\psi^{+}(t_{f}\!)-\psi^{-}(t_{f}\!)]]
\nonumber \\
&& \qquad\qquad \times {\rm exp}\left[\imath \int
\mathrm{d}^{d-1} \vec{x}\int_{t_{0}}^{t_{f}} \mathrm{d}t^\prime
\left({\cal L}[\phi^{+},\psi^{+},t']-{\cal
L}[\phi^{-},\psi^{-},t'] +J_{+}^{\phi}\phi^{+} +
J_{-}^{\phi}\phi^{-} + [J_{+}^{\psi} \psi^{+}] +
[J_{-}^{\psi}\psi^{-}] \right)\right]  
\end{eqnarray}
where for notational convenience we have written $[{\cal D}\psi]={\cal D}\bar{\psi}{\cal D}\psi$, $[J_{+}^{\psi} \psi^{+}]= J_{+}^{\psi} \psi^{+}+\bar{\psi}^{+}\bar{J_+}^{\bar\psi^+} $  etc. 
   The fields $\phi,\, \psi,\,\bar{\psi} $ have their sources $J^{\phi}$,
 $J^{\psi}$ $\bar{J}^{\bar\psi}$ respectively. Note that there are two species of fields and sources for each category. The $+$ sign denotes forward evolution in time whereas the $-$ sign denotes backward evolution. The $\delta$-function ensures the field configurations are the same on the final hypersurface at $t=t_f$. The expectation values of $n$-point functions can be found by taking functional differentiation of the generating functional~\ref{Z:inin} with respect to the sources $J$'s, for instance
\begin{eqnarray} \label{npointfunctions}
\left. \mathrm{Tr}\left[ {\rho}(t_{0})
\overline{T}[\phi(x_1)\dots \phi(x_n)]
T[\phi(y_1)\dots \phi(y_k)] \right] =
\frac{\delta^{n+k}{\cal Z}[J, \rho(t_0)] } {\imath\delta \!
J_{-}^{\phi}(x_{1})\cdots \imath\delta\! J_{-}^{\phi}(x_{n})
\imath\delta\! J_{+}^{\phi}(y_{1})\cdots \imath\delta\!
J_{+}^{\phi}(y_{k})} \right|_{J = 0}  
\end{eqnarray}
%

With the help of the above, we now define the following propagators for the scalar field,
\begin{subequations}
\label{propagators}
\begin{eqnarray}
\imath\Delta^{++}_{\phi}(x,x^\prime) &=&
\mathrm{Tr}\left[{\rho}(t_{0})
T[\phi(x^\prime)\phi(x)] \right] =
 \mathrm{Tr}\left[{\rho}(t_{0}) \phi^+(x)\phi^+(x^\prime)\right] =
\left.\frac{\delta^2{\cal Z}[J, \rho(t_0)]} {\imath\delta\!
J_{+}^{\phi}(x) \imath\delta\! J_{+}^{\phi}(x^\prime)}
\right|_{J=0}
\label{propagatorsa} \\
\imath\Delta^{--}_{\phi}(x,x^\prime) &=&
 \mathrm{Tr}\left[{\rho}(t_{0}) \overline{T} [ \phi(x^\prime)\phi(x)]
\right] =
 \mathrm{Tr}\left[{\rho}(t_{0}) \phi^-(x^\prime)\phi^-(x)\right] =
\left. \frac{\delta^2{\cal Z} [J, \rho(t_0)]} {\imath\delta\!
J_{-}^{\phi}(x)\imath\delta\! J_{-}^{\phi}(x^\prime)}
\right|_{J=0}
\label{propagatorsb} \\
\imath\Delta^{-+}_{\phi}(x,x^\prime) &=&
 \mathrm{Tr}\left[{\rho}(t_{0})  \phi(x)\phi(x^\prime)\right] =
 \mathrm{Tr}\left[{\rho}(t_{0}) \phi^-(x)\phi^+(x^\prime)\right] =
\left.\frac{\delta^2{\cal Z} [J, \rho(t_0)] } {\imath\delta\!
J_{-}^{\phi}(x)\imath\delta\! J_{+}^{\phi}(x^\prime)}
\right|_{J=0} \label{propagatorsc}
\\
\imath\Delta^{+-}_{\phi}(x,x^\prime) &=&
 \mathrm{Tr}\left[{\rho}(t_{0})\phi(x^\prime)\phi(x)\right] =
 \mathrm{Tr}\left[{\rho}(t_{0})\phi^-(x^\prime)\phi^+(x)\right]=
\left.\frac{\delta^2{\cal Z} [J, \rho(t_0)] } {\imath\delta\!
J_{+}^{\phi}(x)\imath\delta\! J_{-}^{\phi}(x^\prime)}
\right|_{J=0}
\label{propagatorsd}
\end{eqnarray}
\end{subequations}
where we have taken $t>t'$ above.
The time ordered and the anti-time ordered propagators are respectively the Feynman and anti-Feynman propagators and the rest are the two Wightman functions. One can also write
\begin{eqnarray}
\label{propagatoridentities}
\imath\Delta^{++}_{\phi}(x,x^\prime) &=& \theta(t-t^\prime)\imath
\Delta^{-+}_{\phi}(x,x^\prime) + \theta(t^\prime-t)\imath
\Delta^{+-}_{\phi}(x,x^\prime) \nonumber
\\
\imath\Delta^{--}_{\phi}(x,x^\prime) &=& \theta(t^\prime-t)\imath
\Delta^{-+}_{\phi}(x,x^\prime) + \theta(t-t^\prime)\imath
\Delta^{+-}_{\phi}(x,x^\prime) 
\end{eqnarray}
These propagators and  Wightman functions also satisfy the following properties, 
\begin{eqnarray}
\imath\Delta^{++}_{\phi}(x,x^\prime) +
\imath\Delta^{--}_{\phi}(x,x^\prime) &=& \imath
\Delta^{-+}_{\phi}(x,x^\prime) +\imath
\Delta^{+-}_{\phi}(x,x^\prime) \label{propagatoridentitiesc}
\\
\imath \Delta^{-+}_{\phi}(x,x^\prime)&=&\imath
\Delta^{+-}_{\phi}(x^\prime,x) \label{propagatoridentitiesd}
\end{eqnarray}
The retarded  and the advanced
 propagators, useful for our later purpose, are respectively defined as
\begin{subequations}
\label{propagators2}
\begin{eqnarray}
\imath\Delta^{\mathrm{r}}_{\phi}(x,x^\prime) &=& \imath
\Delta^{++}_{\phi}(x,x^\prime) - \imath
\Delta^{+-}_{\phi}(x,x^\prime)
\\
\imath\Delta^{\mathrm{a}}_{\phi}(x,x^\prime)&=& \imath
\Delta^{++}_{\phi}(x,x^\prime) - \imath
\Delta^{-+}_{\phi}(x,x^\prime)
\label{Delta:adv}
\end{eqnarray}
\end{subequations}
One also defines the spectral two point function and the statistical propagator as (e.g.~\cite{koksma}),
\begin{eqnarray} \label{Delta:causal}
&&\imath\Delta^{c}_{\phi} (x,x^\prime) = \mathrm{Tr}\left(
{\rho}(t_{0})  [\phi(x),\phi(x^\prime)]\right)= \imath
\Delta^{-+}_{\phi}(x,x^\prime) - \imath
\Delta^{+-}_{\phi}(x,x^\prime)\nonumber\\
&&F_{\phi}(x,x') = \frac{1}{2} \mathrm{Tr}\left[ {\rho}(t_{0})
[ \phi(x),\phi(x')]_+ \right]=
\frac{1}{2}\Big(\imath\Delta^{-+}_{\phi}(x,x') +
\imath\Delta^{+-}_{\phi}(x,x')\Big) 
\end{eqnarray}
The spectral function can tell us about the states of the system and the spectrum, but  it does not hold any information on how these states are occupied. The statistical propagator on the other hand, gives information about how the states are populated.  Hence the latter is  much relevant  to study decoherence and entropy~\cite{NEQFT} (also references therein).

%
%
The free Feynman, anti-Feynman and the Wightman  functions  satisfy
\begin{equation} \label{Feynman propagator}
(\partial_x^2 - m^2)\imath \Delta^{ss'}_{\phi}(x,x^\prime) =
\imath s \delta_{ss'}\delta^{\scriptscriptstyle{D}}\!(x-x^\prime)\qquad (s,s'=\pm,\,\,\,\,{\rm no~sum~on~}s)
\end{equation}

The propagators for the fermionic field can be defined in a likewise manner, keeping in mind the anti-commutations satisfied by them and the replacement of the d'Alembertian $\partial^2$ by $\slashed{\partial}$. We shall consider a zero temperature field theory, and hence various averages such as in \ref{propagators} are to be understood simply as the vacuum expectation values below.

With these equipments, we are now ready to compute the loop correction to the two-point functions and to find out the Kadanoff-Baym equation, i.e. the equation of motion satisfied by them.

\section{Derivation of the Kadanoff-Baym equations}
\label{The Kadanoff-Baym equations}

The Kadanoff-Baym equations are integro-differential equations satisfied by the two-point functions in an interacting quantum field theory \cite{KadanoffBaym:1962}. It is well known that the  one particle irreducible effective action, when varied with respect to some background field, gives rise to the quantum corrected field equation. Likewise, the 2PI effective action, when varied with respect to the two point functions, yields equation of motion satisfied by them known as  the Kadanoff-Baym equations. Since the 2PI effective action contains quantum corrections, we obtain extensions of free theory equations like \ref{Feynman propagator}, essentially containing  the effect of loops.  
 One can obtain the 2PI effective action as a double Legendre transform from the generating functional for connected Green functions with respect to the linear source $J$ and also another quadratic source~\cite{NEQFT, Calzetta:1986cq, cornwall, jackiw, 2PIyukawa}. These equations contain the effects of the non-local self energy.  
 
\begin{figure}[t!]
  \centering
   \includegraphics[width=\textwidth]{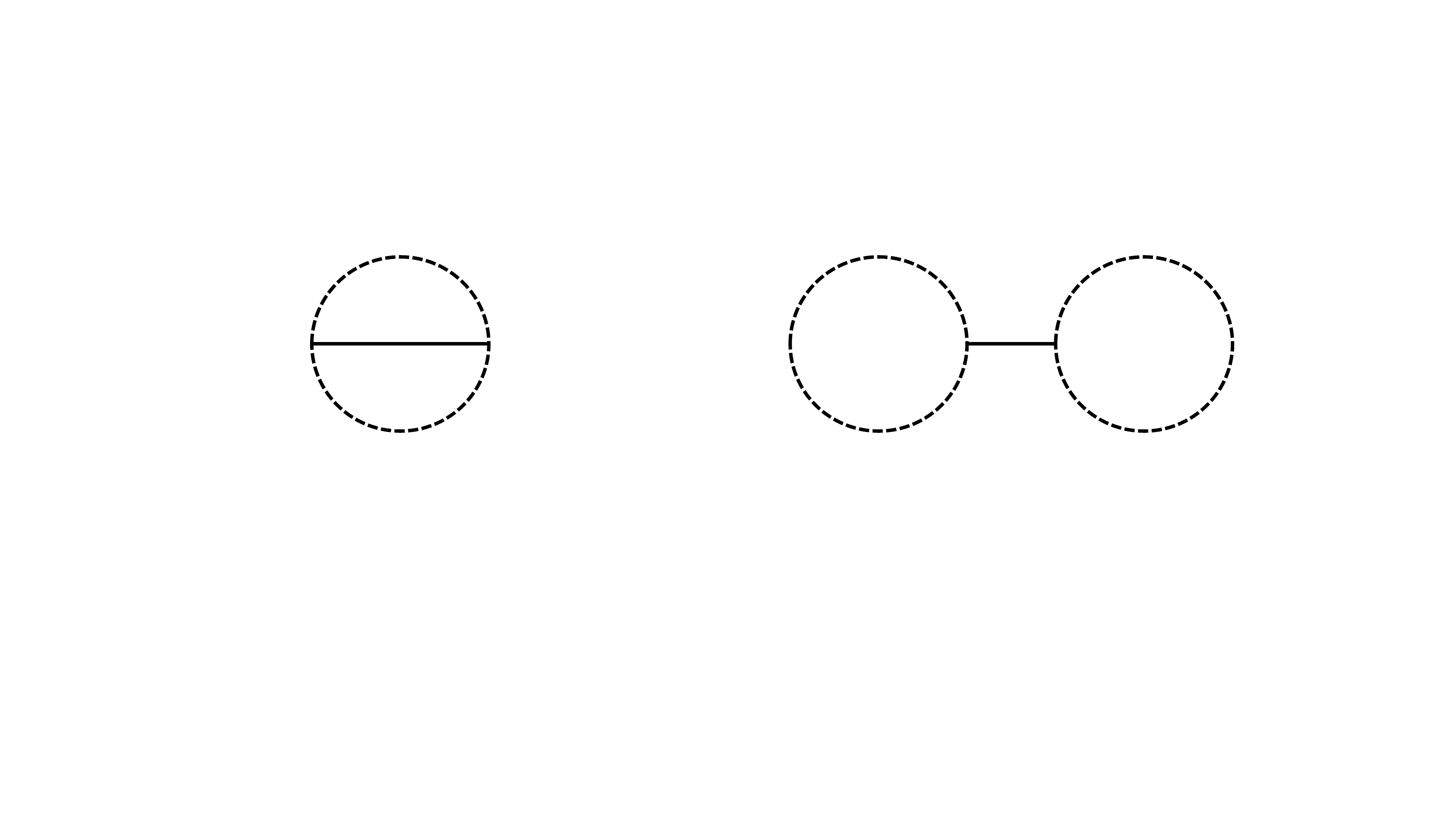}
   \vspace*{-40mm}
   \caption{\small \it Contributions to the 2PI effective action up to two
     loop order. The solid lines denote $\phi$-propagators, whereas the dashed lines correspond
     to $\psi$-propagators. The tadpoles will not make any contribution. See main text for discussion.   \label{fig:2PIEfAction}}
 \end{figure}

 We expand the effective action corresponding to \ref{action:tree1} up to two loop order as
\begin{eqnarray}\label{effectiveaction}
\Gamma[{\phi}^{s},{\psi}^{s},{\bar \psi}^s, \imath\Delta_{\phi}^{ss^{\prime}},\imath{}S^{ss^\prime}_{\psi}]
&&= S[{\phi}^{s},{\psi}^{s}] + \frac{\imath}{2}
\mathrm{Tr} \ln [ (\imath\Delta_{\phi}^{ss^{\prime}})^{-1}] 
{-\imath} \mathrm{Tr} \ln [
(\imath{}S^{ss^\prime}_{\psi})^{-1}] \nonumber\\
&&  +\frac{1}{2} \mathrm{Tr} \frac{\delta^{2}\!
S[{\phi}^{s},{\psi}^{s}]}{\delta\!{\phi}^{s} \delta\!
{\phi}^{s^\prime}} \imath\Delta_{\phi}^{ss^{\prime}} -
\mathrm{Tr} \frac{\delta^{2}
S[{\phi}^{s},{\psi}^{s}]}{\delta\! {\bar \psi}^{s} \delta\!
{\psi}^{s^\prime}} \imath {}S^{ss^\prime}_{\psi} +
\Gamma^{(2)}[{\phi}^{s},{\psi}^{s},\imath\Delta_{\phi}^{ss^{\prime}},\imath {}S^{ss^\prime}_{\psi}]\quad
\end{eqnarray}
where $s,s'=\pm$,  and  $\imath {}S^{ss^\prime}_{\psi}$ are the fermion propagators. $\Gamma^{(2)}$ denotes the 2PI contribution to the effective action at two loop, as shown in  \ref{fig:2PIEfAction}. Note however that since we are considering massless fermions, the tadpoles vanish. Even with a massive fermion, the tadpoles can be completely renormalised away in the flat spacetime. Thus we need to consider  only the sunset like diagram of \ref{fig:2PIEfAction}.
We now explicitly write down the parts of \ref{effectiveaction} containing quantum corrections as
\begin{subequations}
\label{Gamma}
\begin{eqnarray}
\Gamma^{(0)}[\imath\Delta_{\phi}^{ss^{\prime}},\imath {}S^{ss^\prime}_{\psi}) ] &=&
\phantom{+} \int \mathrm{d}^{\scriptscriptstyle{d}}\!x
\mathrm{d}^{\scriptscriptstyle{d}}\!x' \sum_{s,s^\prime=\pm}
\frac{s}{2}(\partial_{x}^{2} -
m^{2})\delta^{\scriptscriptstyle{d}}\!(x-x') \delta^{ss^\prime}
\imath\Delta^{s^{\prime}s}_{\phi}(x',x)\label{Gamma0} \\
&& - \int \mathrm{d}^{\scriptscriptstyle{d}}\! x
\mathrm{d}^{\scriptscriptstyle{d}}\! x' \sum_{s,s^\prime=\pm}
{s}\imath\slashed{\partial}_{x}\delta^{\scriptscriptstyle{d}}\! (x-x') \delta^{ss^\prime}
\imath {}S^{s^{\prime}s}_{\psi} (x,x') \nonumber
\\
\Gamma^{(1)} [\imath\Delta_{\phi}^{ss^{\prime}},\imath{}S^{ss^\prime}_{\psi} ]
&=& -\frac{\imath}{2}{\rm Tr} \ln\left[\imath \Delta^{ss}_{\phi}
(x,x)\right]  +{\imath}{\rm Tr} \ln\left[\imath {}S^{ss}_{\psi} (x,x)\right] \label{Gamma1}
\\
\Gamma^{(2)}[\imath\Delta_{\phi}^{ss^{\prime}},\imath {}S^{ss^\prime}_{\psi} ] &=&
 \frac{\imath
g^{2}}{2} \int \mathrm{d}^{\scriptscriptstyle{d}}\! x
\mathrm{d}^{\scriptscriptstyle{d}}\! x' \sum_{s,s^\prime=\pm}ss^\prime
{\rm Tr}[iS^{ss'}_\psi(x,x')iS^{s's}_\psi(x',x)]
\imath\Delta^{ss^{\prime}}_{\phi}(x,x') \label{Gamma2} ,
\end{eqnarray}
\end{subequations}
where the last line corresponds to the first of \ref{fig:2PIEfAction}.

We now vary the effective action with respect to the bosonic  and the fermionic propagators, $i\Delta_{\phi}$ and $iS_{\psi}$, to obtain the equations of motion
\begin{eqnarray}
\frac{ \delta \Gamma[\imath\Delta_{\phi}^{ss^{\prime}},
\imath {}S^{ss^\prime}_{\psi}
]}{\delta \imath\Delta_{\phi}^{ss^{\prime}}} = 0, \label{EOM1a} \qquad {\rm and}\qquad 
\frac{\delta
\Gamma[\imath\Delta_{\phi}^{ss^{\prime}},\imath {}S^{ss^\prime}_{\psi} ]}{\delta
\imath {}S^{ss^\prime}_{\psi}} = 0 \label{EOM1}
\end{eqnarray}
explicitly giving respectively
\begin{subequations}
\label{EOM2}
\begin{eqnarray}
s(\partial_{x}^{2} -
m^{2})\delta^{\scriptscriptstyle{d}}\!(x-x^\prime)
\delta^{ss^\prime}
-\imath\left[\imath\Delta^{ss^\prime}_{\phi}(x,x^\prime)\right]^{-1}
+ \imath g^{2}ss^\prime
{\rm Tr}[iS^{ss'}_\psi(x,x')iS^{s's}_\psi(x',x)] &=& 0
\label{EOM2a} \\
s\imath\slashed{\partial}_{x}\delta^{\scriptscriptstyle{d}}\!(x-x^\prime)\delta^{ss^\prime}
-\imath\left[\imath {}S^{ss^\prime}_{\psi}(x,x^\prime)\right]^{-1}
 -\imath
g^{2}ss^\prime \, \imath {}S^{s^\prime s}_{\psi}(x',x)
\imath\Delta^{ss^\prime}_{\phi}(x,x') &=& 0 \label{EOM2b} 
\end{eqnarray}
\end{subequations}

We next multiply \ref{EOM2a} and \ref{EOM2b} respectively by $s
\imath\Delta^{s^{\prime}s^{\prime\prime}}_{\phi}(x^\prime,x^{\prime\prime})$ and $s
\imath {}S^{s^{\prime}s^{\prime\prime}}_{\psi}(x^\prime,x^{\prime\prime})$ (from the right),
and then integrate over $x^\prime$ and sum
over $s^\prime=\pm$. Recalling that the integration over the product of the propagator and its inverse gives a $\delta$-function,   we obtain the one-loop Kadanoff-Baym equations 
\begin{subequations}
\label{EOM3}
\begin{eqnarray}
(\partial_{x}^{2}-m^{2})\imath
\Delta^{ss^\prime}_{\phi}(x,x^\prime) -\sum_{s^{\prime\prime}=\pm}s^{\prime\prime}\int
\mathrm{d}^{\scriptscriptstyle{d}}\! x''
i M^{ss^{\prime\prime}}_{\phi}(x,x'')\imath \Delta^{s^{\prime\prime}s^{\prime}}_{\phi}(x'',x^\prime) &=&
\imath s\delta^{ss^\prime}\delta^{\scriptscriptstyle{d}}\!(x-x^\prime)
\label{EOM3a} \\
\imath\slashed{\partial} \imath{}S^{ss^\prime}_{\psi}(x,x^\prime) -\sum_{s^{\prime\prime}=\pm}s^{\prime\prime}\int
\mathrm{d}^{\scriptscriptstyle{d}}\! x''
iM^{ss^{\prime\prime}}_{\psi}(x,x'')\,\imath {}S^{s^{\prime\prime}s^{\prime}}_{\psi}(x'',x^\prime) &=&
\imath s\delta^{ss^\prime}\delta^{\scriptscriptstyle{d}}\!(x-x^\prime)
\label{EOM3b} 
\end{eqnarray}
\end{subequations}
where  we have abbreviated the one loop the self-energies as
\begin{subequations}
\label{selfMass}
\begin{eqnarray}
\imath M^{ss^{\prime\prime}}_{\phi}(x,x') &=&
 -\imath g^{2} {\rm Tr}[iS^{ss'}_\psi(x,x')iS^{s's}_\psi(x',x)] =-2ss^{\prime\prime}\frac{\delta\Gamma^{(2)}[\imath\Delta_{\phi}^{ss^{\prime}},
\imath {}S^{ss^\prime}_{\psi} ]}{\delta\imath\Delta^{s^{\prime\prime}s}_{\phi}}
\label{selfMassa} \\
 \imath M^{ss^{\prime\prime}}_{\psi}(x,x') &=&
   \imath g^{2} \imath {}S^{s^{\prime\prime}s}_{\psi}(x',x) \imath \Delta^{ss^{\prime\prime}}_{\phi}(x,x')=ss^{\prime\prime}\frac{\delta\Gamma^{(2)}[\imath\Delta_{\phi}^{ss^{\prime}},
\imath {}S^{ss^\prime}_{\psi} ]}{\delta\imath {}S^{s^{\prime\prime}s}_{\psi}}
\label{selfMassb} 
\end{eqnarray}
\end{subequations}
Since we are considering the scalar to be our system which is observed, we shall consider only the scalar self energy, \ref{selfMassa}. The corresponding Feynman diagram is given by the first of~\ref{figa}. Accordingly, by expanding the summations, we rewrite \ref{EOM3a} as
\begin{subequations}
\label{EOM3aExtended}
\begin{eqnarray}
(\partial_{x}^{2}-m^{2}) \imath
\Delta^{++}_{\phi}(x,x^\prime) - \int
\mathrm{d}^{\scriptscriptstyle{d}}\! y \left[\imath
M^{++}_{\phi}(x,y)\imath \Delta^{++}_{\phi}(y,x^\prime) - \imath
M^{+-}_{\phi}(x,y)\imath \Delta^{-+}_{\phi}(y,x^\prime)\right] &=& \imath\delta^{\scriptscriptstyle{d}}(x-x^\prime)
\label{EOM3a++} \\
(\partial_{x}^{2} - m^{2})\imath
\Delta^{+-}_{\phi}(x,x^\prime)- \int
\mathrm{d}^{\scriptscriptstyle{d}}\!y \left[\imath
M^{++}_{\phi}(x,y)\imath \Delta^{+-}_{\phi}(y,x^\prime) - \imath
M^{+-}_{\phi}(x,y)\imath \Delta^{--}_{\phi}(y,x^\prime)\right] &=&\,
0
\label{EOM3a+-} \\
(\partial_{x}^{2} - m^{2})\imath
\Delta^{-+}_{\phi}(x,x^\prime) - \int
\mathrm{d}^{\scriptscriptstyle{d}}\!y \left[\imath
M^{-+}_{\phi}(x,y)\imath \Delta^{++}_{\phi}(y,x^\prime) - \imath
M^{--}_{\phi}(x,y)\imath \Delta^{-+}_{\phi}(y,x^\prime)\right] &=&\,
0
\label{EOM3a-+} \\
(\partial_{x}^{2}-m^{2})\imath
\Delta^{--}_{\phi}(x,x^\prime) - \int
\mathrm{d}^{\scriptscriptstyle{d}}\!y \left[\imath
M^{-+}_{\phi}(x,y)\imath \Delta^{+-}_{\phi}(y,x^\prime) - \imath
M^{--}_{\phi}(x,y)\imath \Delta^{--}_{\phi} (y,x^\prime)\right]
&=&-\imath \delta^{\scriptscriptstyle{d}}(x-x^\prime)
\label{EOM3a--}
\end{eqnarray}
\end{subequations}
Setting $g=0$ above makes the self energies vanishing, thereby reproducing the free theory results of~\ref{Feynman propagator}.  Note also that even though the self energies appearing in the above equations are of one loop order, we may integrate these equations to find out the propagators, eventually {\it non-perturbative} in the coupling constant. In other words, the Kadanoff-Baym equations gives a framework to resum the self energies. This seems to be in particular useful in the context of the primordial cosmic inflation, where late time secular effects may be present, necessitating resummation, e.g.~\cite{Cabrer:2007xm,Tsamis:2005hd,Brunier:2004sb}.

We shall solve \ref{EOM3aExtended} by going to the momentum space. We define the Fourier transform
\begin{eqnarray}
\imath \Delta_{\phi}^{ss^{\prime}}(x,x') &=& \int
\frac{\mathrm{d}^{\scriptscriptstyle{d}}k}{(2\pi)^{\scriptscriptstyle{d}}}
 \imath \Delta_{\phi}^{ss^{\prime}}(k){\rm e}^{\imath k \cdot (x-x')}
\label{Fouriertransformdef2a} 
\end{eqnarray}
in terms of which  \ref{EOM3aExtended} become
\begin{subequations}
\label{EOM4Fourier}
\begin{eqnarray}
(-k^2-m^{2}- \imath M^{++}_{\phi}(k))
\imath \Delta^{++}_{\phi}(k) + \imath M^{+-}_{\phi}(k)
\imath \Delta^{-+}_{\phi}(k) &=& \, \imath
\label{EOM4Fourier++} \\
(-k^2 - m^{2} -\imath M^{++}_{\phi}(k)
)\imath \Delta^{+-}_{\phi}(k) + \imath
M^{+-}_{\phi}(k)\imath \Delta^{--}_{\phi}(k) &=& 0
\label{EOM4Fourier+-} \\
(-k^2 - m^{2} + \imath M^{--}_{\phi}(k)
)\imath \Delta^{-+}_{\phi}(k) - \imath
M^{-+}_{\phi}(k)\imath \Delta^{++}_{\phi}(k) &=& 0
\label{EOM4Fourier-+} \\
(-k^2-m^{2} + \imath
M^{--}_{\phi}(k))\imath \Delta^{--}_{\phi}(k) - \imath
M^{-+}_{\phi}(k)\imath \Delta^{+-}_{\phi}(k) &=&
-\imath \label{EOM4Fourier--}
\end{eqnarray}
\end{subequations}
where  $k^2=k_{\mu}k^{\mu}=-k_{0}^{2}+|\vec{k}|^{2}$. Solving these coupled algebraic equations, we can find out various two point functions. For example, on subtracting \ref{EOM4Fourier++} from \ref{EOM4Fourier-+}, we have the momentum space expression for the advanced propagator \ref{Delta:adv}, 
\begin{equation}\
\imath \Delta^{\mathrm{a}}_{\phi}(k) = \imath \Delta^{++}_{\phi}(k)-\imath \Delta^{-+}_{\phi}(k)=  \frac{-\imath}{
k^2+m^{2}+\imath
M^{\mathrm{a}}_{\phi}(k)}
 \label{FourierAdvanced1} 
\end{equation}
where the advanced self-energy $\imath M^{\mathrm{a}}_{\phi}(k)$ is given by
$$\imath M^{\mathrm{a}}_{\phi}(k) = \imath
M^{++}_{\phi}(k) - \imath
M^{-+}_{\phi}(k) =  \imath M^{+-}_{\phi}(k) - \imath
M^{--}_{\phi}(k)$$
Likewise we find the Wightman functions
\begin{subequations}
\label{FourierWightman1}
\begin{eqnarray}
\imath \Delta^{-+}_{\phi}(k) &=& \frac{ - \imath
M^{-+}_{\phi}(k) \imath
\Delta^{\mathrm{a}}_{\phi}(k)}{k^2+m^{2}+\imath
M^{\mathrm{r}}_{\phi}(k)}
 \label{FourierWightmana1} \\
\imath \Delta^{+-}_{\phi}(k) &=& \frac{ - \imath
M^{+-}_{\phi}(k) \imath
\Delta^{\mathrm{a}}_{\phi}(k)}{k^2+m^{2}+\imath
M^{\mathrm{r}}_{\phi}(k)}
\label{FourierWightmanb1} 
\end{eqnarray}
\end{subequations}
where the retarded self-energy $\imath M^{\mathrm{r}}_{\phi}(k)$ is given by
  $$\imath M^{\mathrm{r}}_{\phi}(k) = \imath
M^{++}_{\phi}(k) - \imath
M^{+-}_{\phi}(k) =  \imath
M^{-+}_{\phi}(k) - \imath M^{--}_{\phi}(k)$$
Substituting \ref{FourierAdvanced1} into 
\ref{FourierWightmana1} and \ref{FourierWightmanb1}, we finally obtain the momentum space expression of the statistical propagator \ref{Delta:causal},
\begin{eqnarray}
\label{statprop1}
    F_{\phi}(k)&&=-\frac{ \imath
\Delta^{\mathrm{a}}_{\phi}(k) (\imath
M^{-+}_{\phi}(k)+\imath
M^{+-}_{\phi}(k))}{2(k^2+m^{2}+\imath
M^{\mathrm{r}}_{\phi}(k))} \nonumber\\ &&=\frac{\imath(\imath
M^{-+}_{\phi}(k)+\imath
M^{+-}_{\phi}(k))}{2(\imath
M^{\mathrm{r}}_{\phi}(k)-\imath
M^{\mathrm{a}}_{\phi}(k))}\Bigg(\frac{1}{k^2+m^2+\imath
M^{\mathrm{a}}_{\phi}(k)}-\frac{1}{k^2+m^2+\imath
M^{\mathrm{r}}_{\phi}(k)}\Bigg)
\end{eqnarray}
Recall that the statistical propagator will yield the expression of the phase space area and entropy, \ref{deltaareainphasespace}, \ref{entropy}. Thus in order to compute the statistical propagator, we need to determine various  self-energies, as appearing in \ref{statprop1}. However, note that the above expressions are not renormalised. Hence we shall use the renormalised self energies in \ref{statprop1}, in order to compute the entropy. Also in particular, note that since the self energies are ${\cal O}(g^2)$, the only 
coupling constant dependence of the above expression comes in the denominator of the terms within the parenthesis. Thus the expression for the statistical propagator is actually non-perturbative and contains  the resummed self-energy corresponding to the series of one-loop diagrams  (i.e., the first of \ref{figa}), owing to the Kadanoff-Baym equations.

\subsection{The retarded self-energy and its renormalisation}
\label{Renormalising the Kadanoff-Baym Equations}

In this section, we compute the self-energy $\imath M_{\phi}^{ss^\prime}(x,x')$ and find out the renormalised retarded self-energy,  $\imath M_{\phi,\mathrm{ren}}^{\mathrm{r}}(x,x')= \imath
M_{\phi,\mathrm{ren}}^{++}(x,x') - \imath M_{\phi}^{+-}(x,x')$, to be useful for our future purpose. Due to the subtraction of the two propagators, it is easily done in coordinate space. 
Necessary techniques in order to deal with such coordinate space computations can be seen in, e.g.~\cite{woodard, Miao:2006pn}. \\

The fermion propagator  $\imath {}S^{ss^\prime}_{\psi}(x,x')$ is obtained  by acting $i\slashed{\partial}$ on the massless scalar field propagator,
\begin{equation}\label{Feynmanpropposition2}
\begin{aligned}
\imath {}S^{ss^\prime}_{\psi}(x,x') = i\slashed{\partial}\left[\frac{\Gamma\left(\frac{d}{2}-1\right)}{4 \pi^{\frac{d}{2}}}\left[\Delta x_{ss^\prime}^{2}\left(x, x^{\prime}\right)\right]^{1-\frac{d}{2}}\right]
=-\frac{i\Gamma\left(\frac{d}{2}\right)}{2 \pi^{\frac{d}{2}}} \frac{\gamma^{\mu} (\Delta x_{\mu})_{ss'}}{\left[\Delta x_{ss^\prime}^{2}\left(x, x^{\prime}\right)\right]^{\frac{d}{2}}}
\end{aligned}
\end{equation}
The Poincarr\'e  invariant biscalar distance  functions with appropriate $i\epsilon$ prescription, $\Delta x_{ss^\prime}^{2}(x,x')$, necessary for the in-in formalism are defined as
\begin{subequations}
\label{x}
\begin{eqnarray}
\Delta x_{++}^{2}(x,x') &=& - \left(\left|t - t' \right| - i
\epsilon \right)^{2} + | \vec{x} -
\vec{x}'|^{2} \label{x++}  \\
\Delta x_{+-}^{2}(x,x') &=& - \left( \phantom{|}t - t'\phantom{|}
+ i \epsilon \right)^{2} + | \vec{x} - \vec{x}'|^{2}
\label{x+-} \\
\Delta x_{-+}^{2}(x,x') &=& - \left( \phantom{|} t - t'\phantom{|}
 - i \epsilon \right)^{2} + | \vec{x} - \vec{x}'|^{2}
\label{x-+} \\
\Delta x_{--}^{2}(x,x') &=& - \left(\left|t - t' \right| + i
\epsilon \right)^{2} + | \vec{x} - \vec{x}'|^{2} \label{x--}
\end{eqnarray}
\end{subequations}
The one loop scalar self-energy $\imath M_{\phi}^{++}(x,x')$ is readily found from  \ref{selfMassa} and \ref{Feynmanpropposition2}
\begin{equation}\label{SelfMassPosspace}
\imath M_{\phi}^{++}(x,x')= - 
\frac{\imath g^{2}\Gamma^{2}(\frac{d}{2})}{ \pi^{\scriptscriptstyle{d}}}
\frac{1}{ \Delta x_{++}^{2\scriptscriptstyle{d}-2}(x,x')} 
\end{equation}
Similarly, we can find out the other self-energies $\imath M_{\phi}^{--}(x,x')$, $\imath M_{\phi}^{+-}(x,x')$ and $\imath M_{\phi}^{-+}(x,x')$ using
the suitable $i\epsilon$ prescriptions as given in \ref{x}. We will now identify the divergence of \ref{SelfMassPosspace}. Note first that for an arbitrary exponent $\alpha$, we have
\begin{equation}\label{SelfMassPosspace2}
\frac{1}{\Delta x_{++}^{2 \alpha}(x,x')} = \frac{1}{4(\alpha-1)(\alpha -
\frac{d}{2})}
\partial^{2} \frac{1}{\Delta x_{++}^{2(\alpha-1)}(x,x')}
\end{equation}
Furthermore, we can write
\begin{subequations}
\label{SelfMassPosspace3}
\begin{equation}
\partial^{2} \frac{1}{\Delta x_{++}^{\scriptscriptstyle{d}-2}(x,x')} = \frac{4
\pi^{\frac{d}{2}}}{\Gamma(\frac{d-2}{2})} \imath
\delta^{\scriptscriptstyle{d}} (x-x') \label{SelfMassPosspace3a}
\end{equation}
For the other distance functions of \ref{x}, we have
\begin{eqnarray}
\partial^{2} \frac{1}{\Delta x_{--}^{\scriptscriptstyle{d}-2}(x,x')} &=& - \frac{4
\pi^{\frac{d}{2}}}{\Gamma(\frac{d-2}{2})} \imath
\delta^{\scriptscriptstyle{d}} (x-x')  \label{SelfMassPosspace3b} \\
\partial^{2} \frac{1}{\Delta x_{+-}^{\scriptscriptstyle{d}-2}(x,x')} &=& \,0=\,
\partial^{2} \frac{1}{\Delta x_{-+}^{\scriptscriptstyle{d}-2}(x,x')}  \label{SelfMassPosspace3d}
\end{eqnarray}
\end{subequations}
We now rewrite \ref{SelfMassPosspace} using
\ref{SelfMassPosspace2} and \ref{SelfMassPosspace3a} as
\begin{equation}\label{SelfMassPosspace4}
\imath M_{\phi}^{++}(x,x')= - \frac{\imath g^{2}
\Gamma^{2}(\frac{d}{2}-1) }{16 \pi^{\scriptscriptstyle{d}}}
\frac{1}{(d-3)(d-4)} \left[ \partial^{4}\left\{ \frac{1}{ \Delta
x_{++}^{2\scriptscriptstyle{d}-6}(x,x')} -
\frac{\mu^{\scriptscriptstyle{d}-4}}{ \Delta
x_{++}^{\scriptscriptstyle{d}-2}(x,x')} \right\} + \frac{ 4
\pi^{\frac{d}{2}} \mu^{\scriptscriptstyle{d}-4}
}{\Gamma(\frac{d-2}{2})} \imath \delta^{\scriptscriptstyle{d}}
(x-x') \right]
\end{equation}
where  $\mu$ is an arbitrary mass scale. We now Taylor expand the terms inside the curly brackets around $d=4$ to obtain,
\begin{equation}\label{SelfTaylor}
\imath M_{\phi}^{++}(x,x')= - \frac{\imath g^{2}
\Gamma(\frac{d}{2}-1) \mu^{d-4} }{4
\pi^{\frac{d}{2}} (d-3)(d-4)}\partial^2 \imath\delta^{\scriptscriptstyle{d}}
(x-x') + \frac{\imath g^{2}}{32 \pi^{4}} \partial^{4}\left[
\frac{\ln(\mu^{2}\Delta x_{++}^{2}(x,x'))}{ \Delta
x_{++}^{2}(x,x')}\right] + \mathcal{O}(d-4) 
\end{equation}

 The first term of the above expression contains an ultraviolet divergence around $d=4$ which we have separated and the second term contains
a non-local contribution to the self-energy.
Since the divergence contains a $\partial^2$, we have to add a scalar field  strength renormalisation counterterm ($\phi \to (1+\delta Z)^{1/2}\phi$).  This yields in the action an additional  kinetic term 
$$\int d^d x\left[ \frac12 \delta Z \phi\, \partial^2  \phi\right]$$
where we have ignored a total divergence. The amputated version of the Feynman diagram corresponding to the above term   involves two functional differentiations with respect to the two scalar field operators, yielding two $\delta$-functions. Integrating either of them, and choosing     
\begin{equation}\label{SelfMasscounterterm}
\delta Z=  \frac{
g^{2} \Gamma(\frac{d}{2}-1) \mu^{d-4} }{4
\pi^{\frac{d}{2}} (d-3)(d-4)},
    \end{equation}
we remove the divergence of \ref{SelfTaylor}. Note that such renormalisation could also be performed at the level of the Kadanoff-Baym equations, \ref{EOM3aExtended}. In that case, the counterterm contribution is achieved by replacing the self energies $\imath M_{\phi}^{++}(x,x')$ and $\imath M_{\phi}^{--}(x,x')$, which contain divergent contributions,  by the amputated counterterm contribution corresponding to $\delta Z$ given above.

$\imath M_{\phi}^{--}(x,x')$ is given by just the complex conjugation of \ref{SelfTaylor}
\begin{equation}\label{SelfMassPosspace4Taylor}
\imath M_{\phi}^{--}(x,x')=  \frac{\imath g^{2}
\Gamma(\frac{d}{2}-1) \mu^{d-4}}{4
\pi^{\frac{d}{2}} (d-3)(d-4)}\partial^2\imath \delta^{\scriptscriptstyle{d}}
(x-x') - \frac{\imath g^{2}}{32 \pi^{4}} \partial^{4}\left[
\frac{\ln(\mu^{2}\Delta x_{--}^{2}(x,x'))}{ \Delta
x_{--}^{2}(x,x')}\right] + \mathcal{O}(d-4) 
\end{equation}
     The divergence appearing in the above expression can be tackled as above, with the counterterm \ref{SelfMasscounterterm}. Thus the  renormalised expressions for the above two self-energies are given by 
\label{SelfMassPosspace5}
\begin{eqnarray}
\imath M_{\phi,\mathrm{ren}}^{++}(x,x')=  \frac{\imath g^{2}}{32 \pi^{4}} \partial^{4}\left[
\frac{\ln(\mu^{2}\Delta x_{++}^{2}(x,x'))}{ \Delta
x_{++}^{2}(x,x')}\right] = (\imath M_{\phi,\mathrm{ren}}^{--}(x,x'))^{\star}
 \label{selfmassa} 
\end{eqnarray}
Also, choosing the suitable pole prescription from \ref{x}, $\imath M_{\phi}^{+-}(x,x')$ and $\imath M_{\phi}^{-+}(x,x')$ are determined by
\begin{eqnarray}
\label{Self5}
\imath M_{\phi}^{+-}(x,x')=   \frac{\imath g^{2}}{32 \pi^{4}} \partial^{4}\left[
\frac{\ln(\mu^{2}\Delta x_{+-}^{2}(x,x'))}{ \Delta
x_{+-}^{2}(x,x')}\right]  = (\imath M_{\phi}^{-+}(x,x'))^{\star}
\end{eqnarray}
Note that $\imath M_{\phi}^{+-}(x,x')$ and $\imath M_{\phi}^{-+}(x,x')$ do not need any renormalisation as they do not 
contain any divergence around $d = 4$, as can be readily verified  from   \ref{SelfMassPosspace3d}.
From \ref{selfmassa} and \ref{Self5},  we now have the renormalised retarded self-energy  
\begin{eqnarray}
\imath M_{\phi,\mathrm{ren}}^{\mathrm{r}}(x,x')= \imath
M_{\phi,\mathrm{ren}}^{++}(x,x') - \imath M_{\phi}^{+-}(x,x') =
\frac{\imath g^{2}}{32 \pi^{4}}
\partial^{4}\left[ \frac{\ln(\mu^{2}\Delta x_{++}^{2}(x,x'))}{
\Delta x_{++}^{2}(x,x')} - \frac{\ln(\mu^{2}\Delta
x_{+-}^{2}(x,x'))}{ \Delta x_{+-}^{2}(x,x')}\right]
\label{R1}
\end{eqnarray}
Using now
\begin{equation}
    \label{prop1}
    \frac{\ln \Delta x^2}{\Delta x^2}=\frac{1}{8} \partial^2 (\ln^2 \Delta x^2-2\ln \Delta x^2)
\end{equation}
we can write the nonlocal terms of \ref{R1} as,
\begin{eqnarray}
\label{prop2}
\frac{\ln \left(\mu^{2} \Delta x_{++}^{2}\right)}{\Delta x_{++}^{2}}-\frac{\ln \left(\mu^{2} \Delta x_{+-}^{2}\right)}{\Delta x_{+-}^{2}}=\frac{\partial^{2}}{8}\left[\ln ^{2}\left(\mu^{2} \Delta x_{++}^{2}\right)-2 \ln \left(\mu^{2} \Delta x_{++}^{2}\right)
-\ln ^{2}\left(\mu^{2} \Delta x_{+-}^{2}\right)+2 \ln \left(\mu^{2} \Delta x_{+-}^{2}\right)\right]
\label{yy}
\end{eqnarray}
Breaking now the logarithms into real and complex parts, we have
\begin{equation}
\label{prop3}
    \ln ^{2}\left(\mu^{2} \Delta x_{++}^{2}\right)-2 \ln \left(\mu^{2} \Delta x_{++}^{2}\right)
-\ln ^{2}\left(\mu^{2} \Delta x_{+-}^{2}\right)+2 \ln \left(\mu^{2} \Delta x_{+-}^{2}\right)=4 \pi i (\ln \mu^2 (\Delta t^2 - \Delta x^2)-1)\theta(\Delta t^2-\Delta x^2)
\end{equation}
Putting these all in together the renormalised retarded self-energy,  \ref{R1}, takes the form
\begin{eqnarray}\label{Re1} 
\imath M_{\phi,\mathrm{ren}}^{\mathrm{r}}(x,x')=\frac{g^{2}}{64 \pi^{3}} \partial^{6}\left[\theta(\Delta
t^2-\Delta x^2)\theta(\Delta t) \left\{1 - \ln\left(\mu^{2}(\Delta
t^{2}-\Delta x^{2})\right)\right\} \right]  
\end{eqnarray}
The step functions  appearing above ensure that the retarded self-energy is non-vanishing only if $\Delta t>0$ and $\Delta
t^2-\Delta x^2>0$. This ensures the expected causal characteristics of the retarded self-energy. 

As we have stated earlier, we wish to make a momentum space computation of the entropy and hence the statistical propagator, \ref{statprop1}.   The relevant renormalised expressions, including that of various self energies are found in  \ref{A} and \ref{B}.   In particular, the  Fourier transform of  \ref{Re1} is done in \ref{B}. Using these ingredients, we wish to compute the von Neumann entropy in the following Section.

\section{Phase space area and entropy}
\label{Phase space area and Entropy}
Renormalised expression for the statistical propagator is given by the renormalised version of~\ref{statprop1}
\begin{eqnarray}
\label{statprop1'}
    F_{\phi}(k)=\frac{\imath(\imath
M^{-+}_{\phi}(k)+\imath
M^{+-}_{\phi}(k))}{2(\imath
M^{\mathrm{r}}_{\phi, {\rm ren}}(k)-\imath
M^{\mathrm{a}}_{\phi, {\rm ren}}(k))}\Bigg(\frac{1}{k^2+m^2+\imath
M^{\mathrm{a}}_{\phi,  {\rm ren}}(k)}-\frac{1}{k^2+m^2+\imath
M^{\mathrm{r}}_{\phi, {\rm ren}}(k)}\Bigg)
\end{eqnarray}
The renormalisation of the various self-energies  has been performed in the preceding Section and \ref{A}, \ref{B}.
Substituting now \ref{Fourierretarded6},
 \ref{FourierWightman2} and \ref{Fourieradvance6} into the above equation, we find after some algebra
\begin{eqnarray}
F_{\phi}(k) &=& - \frac{\imath}{2} \mathrm{sgn}(k^{0})
\theta(k_{0}^{2}-|\vec{k}|^{2})
\Bigg[\frac{1}{k^2+m^{2}+\frac{g^{2}k^2}{8
\pi^{2}}\ln\left(\frac{k^2}{4\mu^{2}}
\right)- \frac{\imath
g^{2}k^2}{8\pi}\mathrm{sgn}(k^{0})\theta((k^{0})^{2}-|\vec{k}|^{2})} \nonumber \\
&& \qquad\qquad\qquad\qquad\qquad -
\frac{1}{k^2+m^{2}+\frac{g^{2}k^2}{8
\pi^{2}}\ln\left(\frac{k^2}{4\mu^{2}}
\right)+ \frac{\imath
g^{2}k^2}{8\pi}\mathrm{sgn}(k^{0})\theta((k^{0})^{2}-|\vec{k}|^{2})} \Bigg]
\label{FourierStatisticals} 
\end{eqnarray}
\begin{figure}[!tbp]
  \centering
  \begin{minipage}[b]{0.45\textwidth}
    \includegraphics[scale=.75]{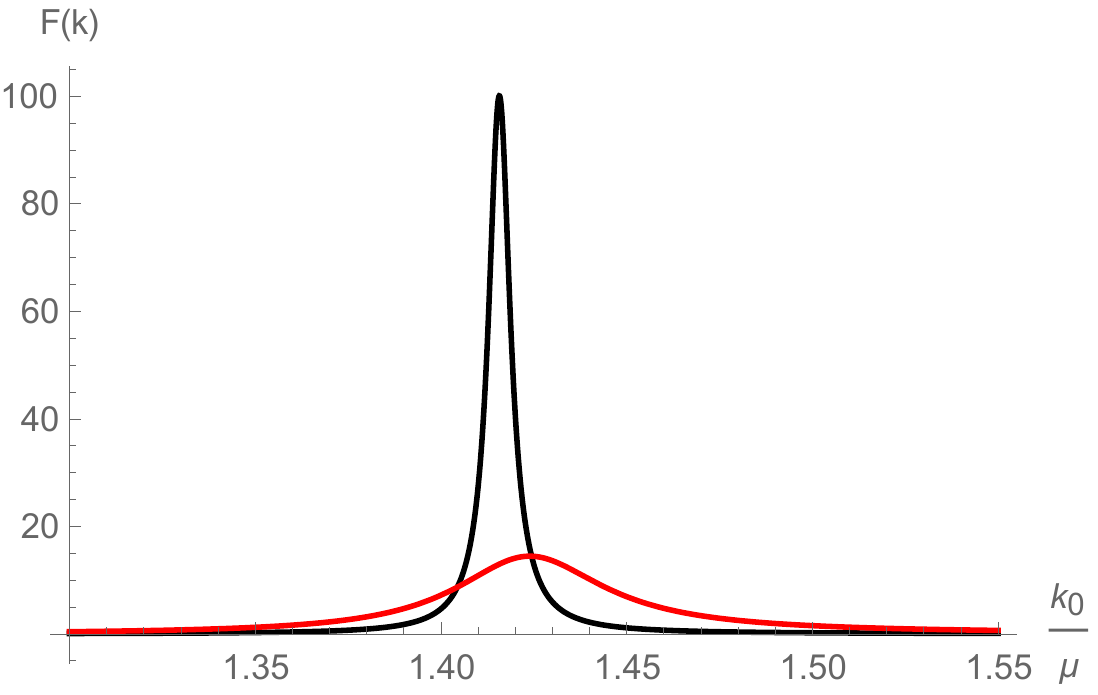}
    \caption{\small \it Variation of the statistical propagator, \ref{FourierStatisticals}, with respect to the dimensionless variable $k^0/\mu$, where we have taken $|\vec{k}|/\mu=1$, $m/\mu=1$ and the Yukawa coupling strengths $g=0.5$ (black curve), $g=1.3$ (red curve).}
     \label{fig:SP}
  \end{minipage}
  \hfill
  \begin{minipage}[b]{0.45\textwidth}
    \includegraphics[scale=.75]{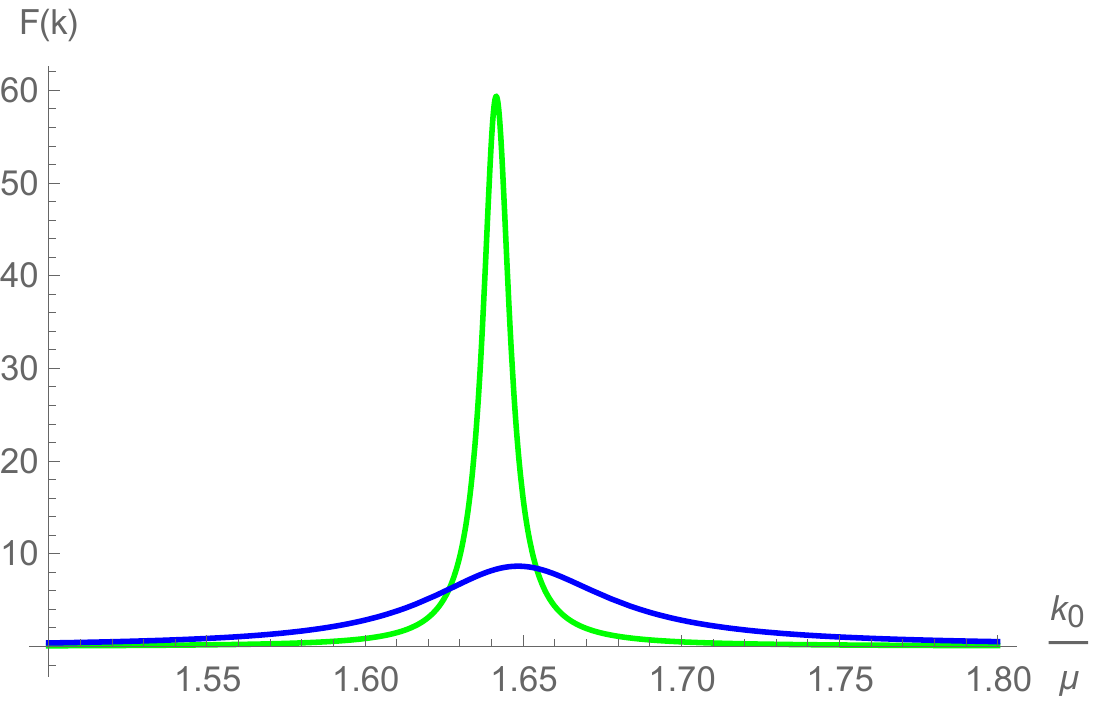}
    \caption{  \small \it Variation of the statistical propagator, \ref{FourierStatisticals}, with respect to the dimensionless variable $k^{0}/\mu$, where we have taken $|\vec{k}|/\mu=1$, $m/\mu=1.3$ and the Yukawa coupling strengths $g=0.5$ (green curve), $g=1.3$ (blue curve). Comparison with \ref{fig:SP} shows lesser numerical value of the statistical propagator with increasing $m/\mu$. See main text for discussion.}
    \label{fig:SP1}
  \end{minipage}
\end{figure}
We have plotted the above statistical propagator with respect to the dimensionless  variable $k^0/\mu$ in~\ref{fig:SP} and \ref{fig:SP1}.
Note that as the Yukawa coupling $g$ gets smaller, $F_{\phi}(k)$ approaches a $\delta$-function dispersion, as expected from \ref{FourierStatisticals}.  However as $g$ increases, the $\delta$-function peak becomes broadened to a quasi-particle peak like that of the  Breit-Wigner kind. Further broadening of the peak with increasing $g$ implies that the resonance becomes broadened and we can no longer sensibly talk about a quasi-particle. Also, these plots show that with the increase in $m/\mu$, the peak gets shifted towards the higher values of $k^0/\mu$ with a decrease in the value of the statistical propagator, which means that the corresponding state becomes less populated.

Now in order to compute the entropy, we would use the definition of the same in terms of the phase space area and the three momentum, \ref{deltaareainphasespace}, \ref{entropy}. We shall assume the  mass of the scalar field to be time independent. However, one can also consider  time-dependent mass as a signature of a non-equilibrium system, as has been  considered in \cite{koksma}. A time dependent mass function will break the time translation invariance, inducing an explicit proper time dependence on the statistical propagator. We also note that in standard quantum field theories, one generally talks about the early or late times when the system is in equilibrium and it is found in some eigenstates of the free Hamiltonian. For a system which is out-of-equilibrium however, an exact distinction between energy states is unclear. A standard approach in such scenario is the adiabatic approximation, in which one specifies a reference  set of approximate states under the assumption of a slowly varying dynamical background. Using the projection of system's evolution onto these approximate states, one may hope to study its dynamics at intermediate times. There are schemes to truncate the adiabatic expansion of the system to form the aforementioned approximate basis set at intermediate times. For example in \cite{koksma, kok}, the Bogoliubov transformations are used to achieve the same.\par 

For turning on the perturbation non-adiabatically which we have not considered here, we consider replacing the coupling constant $g$ with $g\to g\,\theta(t-t_0)$. The step function changes the limit of integration \ref{EOM3aExtended} from $-\infty$ to $t_0$ and $t_0$ to some final time. The self energies vanish in the absence of interaction and would only contribute for the   second time interval.  Thus it is clear that the Schwinger-Keldysh contours presented in \ref{fig:schwingercontour1} are not equivalent to this abrupt switching case. Note that turning on the interaction abruptly may certainly lead to particle creation, which may significantly modify the statistical propagator. Some relevant discussion on this can be seen in eg. \cite{koksma}.

\begin{figure}
    \centering
    \includegraphics[scale=0.6]{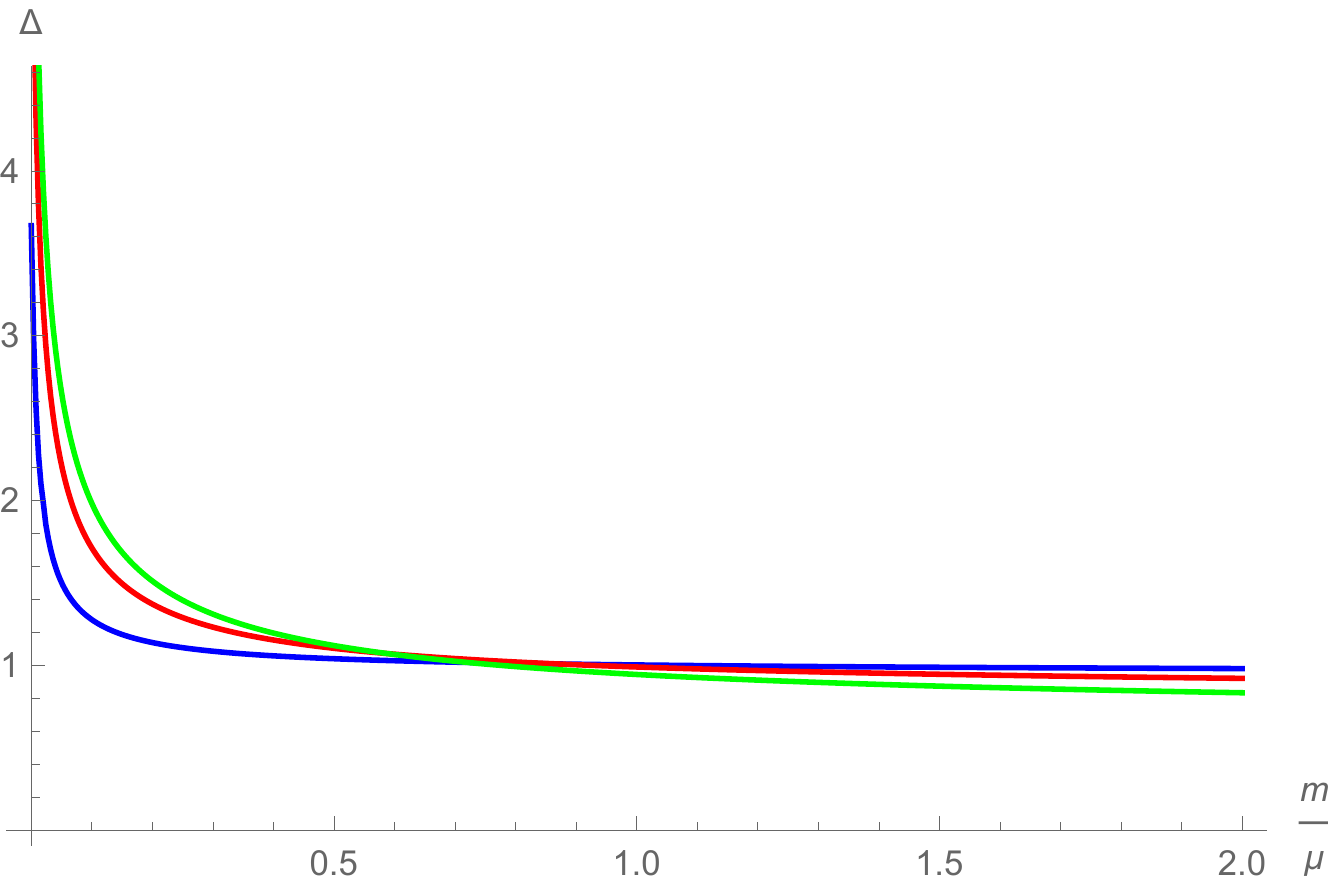}
    \caption{\small \it Variation of the dimensionless phase space area $\Delta$, with respect to the dimensionless mass parameter $m/\mu$ of the system, where we have taken $|\vec{k}|/\mu=1$ and, for $g=0.5$ (blue curve), $g=1$ (red curve) and $g=1.5$ (green curve). See main text for discussion.}
    \label{fig:areamass}
\end{figure}

Using now the Fourier transforms, we have from \ref{FourierStatisticals}
\begin{subequations}
\label{Fconstantmass2}
\begin{eqnarray}
F_{\phi}(|\vec{k}|,0) &=& \int_{-\infty}^{\infty} \frac{\mathrm{d}k^{0}}{2\pi} F_{\phi}(k) \label{Fconstantmass2a} \\
\left.\partial_{t} F_{\phi}(|\vec{k}|, \Delta t) \right|_{\Delta t =0} &=&
- \imath \int_{-\infty}^{\infty} \frac{\mathrm{d}k^{0}}{2\pi}
k^{0}
F_{\phi}(k) \label{Fconstantmass2b} \\
\left.\partial_{t'}\partial_{t} F_{\phi}(|\vec{k}|, \Delta t)
\right|_{\Delta t =0} &=& \int_{-\infty}^{\infty}
\frac{\mathrm{d}k^{0}}{2\pi} (k^{0})^{2} F_{\phi}(k)
\label{Fconstantmass2c}
\end{eqnarray}
\end{subequations}
Substituting now \ref{FourierStatisticals} into the above integrals, we have 
evaluated them numerically. For example, for $|\vec{k}|/\mu=1$, $m/\mu=2$ and $g=0.5$, we
find the numerical value of the phase space area \ref{deltaareainphasespace}, to be
\begin{equation}
\Delta \approx 2.69,  \label{DeltaconstantMass} 
\end{equation}
which is indeed greater than unity and hence indicating a non-vanishing von Neumann entropy, as dictated by~\ref{entropy}.
We have further analysed the variation of the phase space area  with respect to the dimensionless system mass  $m/\mu$, by fixing all the other parameters in \ref{fig:areamass}. Thus $\Delta$  decreases monotonically with increasing $m/\mu$ and asymptotically reaches unity, indicating very small or almost vanishing entropy. This corresponds to the fact   that with the increasing mass, the system becomes more stable, i.e., it becomes difficult for the surrounding to disturb a heavier system.

From \ref{entropy}, we obtain the non-vanishing von Neumann  entropy corresponding to the phase space area of \ref{DeltaconstantMass} ($|\vec{k}|/\mu=1$, $m/\mu=2$ and $g=0.5$)
\begin{equation}
S \approx 1.27\label{SconstantMass} 
\end{equation}
\begin{figure}[!ht]
    \centering
    \includegraphics[scale=0.53]{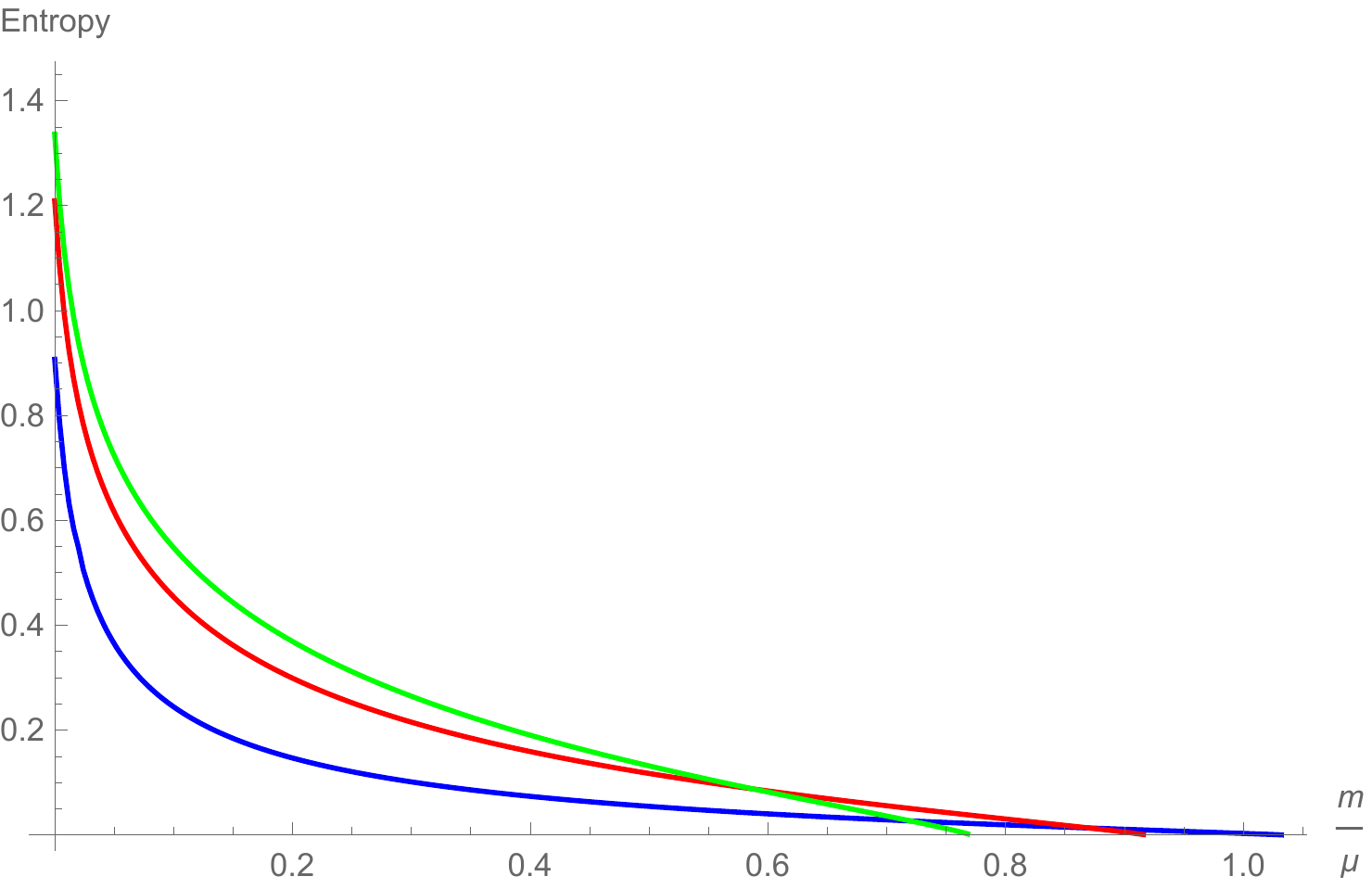}
    \caption{\small \it Variation of the von Neumann entropy with respect to the dimensionless mass parameter $m/\mu$ of the system. We have taken $|\vec{k}|/\mu=1$ and, $g=0.5$ (blue curve), $g=1$ (red curve) and $g=1.5$ (green curve). As of the phase space area, \ref{fig:areamass}, the entropy also decreases monotonically with increasing $m/\mu$. See main text for discussion.   }
    \label{fig:Sconstantmass}
\end{figure}
We have also analysed the variation of the entropy with respect to the system mass by fixing all the other parameters in \ref{fig:Sconstantmass}. Due to the aforementioned reason as that of the phase space area, the entropy is decreasing with the increasing $m/\mu$. 

As we have emphasised earlier, such non-vanishing entropy is due to the
coupling between the system and the environment along with the ignorance of all  kinds of correlations between them for an observer located in the practical world.  The non-trivial statistical  propagator  gives up phase space for the system field that previously was
inaccessible to it. Such increase in the  accessible phase space for the system in turn, implies that less information about the system
field is accessible to the observer, and hence we observe a non-vanishing von Neumann entropy. Due to this reason, as our results show, the increase in the coupling strength increases the von Neumann entropy.

\section{Conclusion}
\label{Conclusion}

This work studies the decoherence  and subsequent entropy generation using the non-equilibrium effective field theory for the Yukawa interaction in the Minkowski spacetime, using the correlator approach of~\cite{JFKTPMGS, koksma, kok}. The scalar is treated as the system, whereas the fermions  as the surrounding and it is assumed that the surrounding or the environment is in its vacuum state. We have assumed as the simplest realistic scenario that the observer measures only the leading and next to the leading  order, two point correlators for the system. The ignorance about all the other correlators yields the generation of entropy.   

We have constructed the Kadanoff-Baym equations from the 2-loop 2PI effective action, i.e. the quantum corrected equations satisfied by the two point correlators in the in-in formalism in~\ref{The Kadanoff-Baym equations}. In \ref{Renormalising the Kadanoff-Baym Equations}, we renormalise our results and finally compute the phase space area, the statistical propagtor and the entropy in \ref{Phase space area and Entropy}. Note that all the results found in the preceding Section are explicitly independent of time, owing to the fact that we have not taken any time dependent background field or time dependent mass. Discussion on the latter scenario can be seen in~\cite{koksma}.

 For a fixed value of all the relevant parameters such as mass, coupling, momentum and energy, we observe a higher numerical value of phase space area and entropy in (cf., \ref{Phase space area and Entropy}), compared to the scenario when both the system and the surrounding are scalars~\cite{koksma}. This is due to the fact that the Yukawa coupling is dimensionless whereas the cubic coupling considered in~\cite{koksma} is dimensionfull, leading to different momentum dependence in the statistical propagator~\ref{FourierStatisticals}. Although we note that the qualitative aspects of the variation  of the entropy (e.g.~\ref{fig:Sconstantmass}) for both the cases are similar.

  Finally, we note that the entire  above analysis might seem to rest upon the {\it implicit} assumption that the observer cannot measure the correlations corresponding to the fermions. For the case when the fermions represent a thermal bath and the scalar is initially at zero temperature, such distinction between the system and the environment seems obvious. For the zero temperature case  we have considered presently, however, such distinction might seem albeit arbitrary. For example, we do not have any {\it explicit} hierarchy of scales of physical quantities here which decides such partitioning. Moreover, we could also have computed the correllators corresponding to the fermions instead. Thus for the zero temperature field theory, the above analysis is based upon the fact that the observer only measures the correlations for the scalar field, and the effect of fermions comes only as virtual particles inside a loop. Such implicit assumption is justified only when the system is much `small' compared to the surrounding, i.e., when it can be considered as a `bath' at zero temperature. This  seems to have some qualitative similarity with the standard formalism of tracing out the fermionic degrees of freedom and look into the effective scalar field dynamics. Now, since we are ignoring a part of the theory, it would lead to an entropy, as we have obtained above. An obvious extension of this would be to go to the two loop self energies of~\ref{figa}, which contains some effect of system's backreaction onto the surrounding. Perhaps more importantly, one should also consider the three-point scalar-fermion-anti-fermion correlators containing the effect of the decay of the scalar. This would  involve constructing an effective action generating those correlators. We reserve these works as future tasks.

As we have emphasised earlier, understanding such decoherence mechanism can be very important in the early inflationary universe scenario, which is in fact our chief motivation. Even though this also must be a zero-temperature case, the short and the super-Hubble wavelength parts of the scalar field might give us some clue to naturally identify the system and the surrounding.
Some related works, using formalism different from this paper can be seen in~\cite{Hollowood:2017bil, Boyanovsky:2018soy, Friedrich:2019hev}. Most importantly, for a massless minimal scalar in an inflationary background, we may expect the appearance of late time, non-perturbative secular effects. Resumming them non-perturbatively using the Kadanoff-Baym equations could be a challenging task.   We hope to return to these issues in our future publications.

\section*{Acknowledgements}
The authors would like to thank anonymous referee for a careful critical reading of the manuscript and for making various useful comments.

\bigskip\bigskip

\appendix
\labelformat{section}{Appendix #1} 
\section{Calculation of renormalised $\imath M_{\phi,{\rm ren}}^{++}(k)$ and $\imath M_{\phi,{\rm ren}}^{--}(k)$}
\label{A}

In this appendix, we compute the renormalised self-energies $\imath M_{\phi}^{++}$ and $\imath M_{\phi}^{--}$
in Fourier space, to be useful for our future purpose. From \ref{selfMassa}, we have in momentum space
\begin{eqnarray}\label{SelfMassFourier1}
\imath M_{\phi}^{++}(k) 
&=& \imath g^{2} {\rm Tr} \int
\frac{\mathrm{d}^{\scriptscriptstyle{d}}p}{(2\pi)^{\scriptscriptstyle{d}}}\Big[
\frac{ \slashed{p}}{p^2-\imath \epsilon} \times
\frac{(\slashed{p}+\slashed{k})}{(p+k)^2-\imath \epsilon}\Big]\nonumber \\
&=& 4\imath g^{2} \int
\frac{\mathrm{d}^{\scriptscriptstyle{d}}p}{(2\pi)^{\scriptscriptstyle{d}}}
\frac{1}{p^2-\imath \epsilon}
\frac{(p^2+k\cdot p)}{(p+k)^2-\imath \epsilon}\nonumber \\
&=& 4\imath g^{2} \int_{0}^{1} \mathrm{d}x \int
\frac{\mathrm{d}^{\scriptscriptstyle{d}}q}{(2\pi)^{\scriptscriptstyle{d}}}
\frac{q^2-k^2x(1-x)}{(q^2+k^2x(1-x)-\imath
\epsilon)^{2}}\nonumber 
\end{eqnarray}
where we have  used Feynman's trick (e.g.~\cite{Peskin:1995ev}). Here $p$ and $k$ are the internal and external momentum corresponding to fermion and scalar propagators respectively.  We have also used
$q=p+xk$, as the new integration
variable. Performing now the usual Euclideanisation of $q$ in the last integral of the above equation, we have
\begin{eqnarray}\label{rr}
\imath M_{\phi}^{++}(k)=-4g^{2} \int_{0}^{1} \mathrm{d}x \int\frac{\mathrm{d}^{\scriptscriptstyle{d}}q_E}{(2\pi)^{\scriptscriptstyle{d}}}
\frac{q_{E}^{2}-k^{2}x(1-x)}{(q_{E}^{2}+k^{2}x(1-x)-\imath\epsilon)^{2}}
\end{eqnarray}
The integral can now straightforwardly be performed,
 yielding
\begin{eqnarray}\label{SelfMassFourier2}
\imath M_{\phi}^{++}(k)&=& -\frac{g^{2}\mu^{d-4}k^2}{4 \pi^{2}}\Big(\frac{1}{d-4}+\frac{\gamma_E}{2}+\ln \sqrt{\pi}\Big)+
\frac{g^{2}k^2}{8\pi^{2}}
\ln\left(\frac{k^2-\imath\epsilon}{4
\mu^{2}}\right) + \mathcal{O}(d-4)
\end{eqnarray}
As of the coordinate space expression \ref{SelfMassPosspace4Taylor}, the divergence can be absorbed in the scalar field strength renormalisation.   Accordingly we have 
\begin{equation}\label{renM++}
\imath M_{\phi,{\rm ren}}^{++}(k)= 
\frac{g^{2}k^2}{8\pi^{2}}
\ln\left(\frac{k^2-\imath\epsilon}{4
\mu^{2}}\right) + \mathcal{O}(d-4)
\end{equation}
$\imath M_{\phi}^{--}(k)$ on the other hand, will be the negative  
of the complex conjugation of the above, follows directly from \ref{selfmassa}, 
\begin{equation}\label{renM--}
\imath M_{\phi,{\rm ren}}^{--}(k)= -
\frac{g^{2}k^2}{8\pi^{2}}
\ln\left(\frac{k^2+\imath\epsilon}{4
\mu^{2}}\right) + \mathcal{O}(d-4)
\end{equation}
We note that the above expression is similar to that  of the case when the environment is also a scalar \cite{koksma}, however in our  case we have one extra factor of $k^2$ multiplied with the logarithm.

\section{Computations for $\imath M_{\phi,\mathrm{ren}}^{\mathrm{r}}(k)$, $\imath M_{\phi,\mathrm{ren}}^{\mathrm{a}}(k)$,  $\imath M^{+-}_{\phi}(k)$ and $\imath M^{-+}_{\phi}(k)$   }\label{B}

In this appendix, we wish to find out the momentum space renormalised expressions of the retarded and advanced self energies as well as  the self energies corresponding to the Wightman functions $\imath M^{+-}_{\phi}(k)$ and $\imath M^{-+}_{\phi}(k)$, as dictated by \ref{statprop1}.

The renormalised expression for the retarded self-energy in coordinate space was found in \ref{Re1}. We shall now take the Fourier transform of it. We do it in two steps, for the sake of convenience of calculations. We first take the Fourier transformation of \ref{Re1} with respect to its spatial part  only
\begin{eqnarray}\label{RetardedSelfMass2}
\imath M_{\phi,\mathrm{ren}}^{\mathrm{r}}(|\vec{k}|, t, t') &=&
\frac{g^{2}}{64 \pi^{3}} (\partial^{2}_{t}+|\vec{k}|^{2})^{3} \int
\mathrm{d}^{3}\Delta{\vec x}\,\, \theta(\Delta
t^2-\Delta x^2)\,\theta(\Delta t) \left[1 - \ln\left(\mu^{2}(\Delta
t^{2}-\Delta x^2)\right) \right] e^{-\imath \vec{k}\cdot\Delta\vec{x}}\nonumber\\
&=& \frac{g^{2}}{16 \pi^{2} |\vec{k}| } (\partial^{2}_{t}+|\vec{k}|^{2})^{3}
\theta(\Delta t) \Delta t^{2} \Bigg[  \frac{\sin(|\vec{k}| \Delta t)-|\vec{k}|
\Delta t \cos(|\vec{k}| \Delta t)}{(|\vec{k}| \Delta
t)^{2}}\left(1-\ln(\mu^{2}\Delta
t^{2}) \right) \nonumber \\
&& \qquad\qquad\qquad\qquad\qquad\qquad - \int_{0}^{1} \mathrm{d}z
\, z \sin(|\vec{k}| \Delta t z)\ln\left(1-z^{2}\right) \Bigg] 
\end{eqnarray}
 The last integral is given by the special function~\cite{AS},
\begin{eqnarray}\label{eu}
&&\xi(|\vec{k}| \Delta t) = \int_{0}^{1} d z z \sin (|\vec{k}| \Delta t z) \ln \left(1-z^{2}\right)\nonumber\\
&&= \frac{2}{(|\vec{k}| \Delta t)^{2}} \sin (|\vec{k}| \Delta t)-\frac{1}{(|\vec{k}| \Delta t)^{2}}[\cos (|\vec{k}| \Delta t)+|\vec{k}| \Delta t \sin (|\vec{k}| \Delta t)]\left[\operatorname{si}(2 |\vec{k}| \Delta t)+\frac{\pi}{2}\right]\nonumber\\&&+\frac{1}{(|\vec{k}| \Delta t)^{2}}[\sin (|\vec{k}| \Delta t)-(|\vec{k}| \Delta t) \cos(|\vec{k}| \Delta t)]\left[\operatorname{ci}(2 |\vec{k}| \Delta t)-\gamma_E-\ln \left(\frac{(|\vec{k}| \Delta t)}{2}\right)\right]
\end{eqnarray}
where $\operatorname{si}(x)$ and $\operatorname{ci}(x)$ are the sine and cosine integral functions, given by \cite{AS}
\begin{eqnarray}\label{sici}
\begin{aligned}
\operatorname{si}(x) &=-\int_{x}^{\infty} d t \frac{\sin t}{t}=-\frac{\pi}{2}+\int_{0}^{x} d t \frac{\sin t}{t} \\
\operatorname{ci}(x) &=-\int_{x}^{\infty} d t \frac{\cos t}{t}=\gamma_E+\ln x+\int_{0}^{x} d t\left[\frac{\cos t-1}{t}\right]
\end{aligned}
\end{eqnarray}
Using  \ref{eu} into \ref{RetardedSelfMass2}, we arrive at 
\begin{eqnarray}\label{RetardedSelfMass3}
\imath M_{\phi,\mathrm{ren}}^{\mathrm{r}}(|\vec{k}|, t, t') &=&
\frac{g^{2}}{16  \pi^{2} |\vec{k}|^{3}} (\partial^{2}_{t}+|\vec{k}|^{2})^{3}
\theta(\Delta t) \Bigg[  \left( |\vec{k}| \Delta t \cos(|\vec{k}| \Delta t) -
\sin(|\vec{k}| \Delta t)\right) \left(\mathrm{ci}(2|\vec{k}|\Delta t) - \gamma_{E} - \ln\left(\frac{|\vec{k}|}{2\mu^{2}\Delta t}\right) -1 \right) \nonumber \\
&& \qquad\qquad\qquad\qquad\qquad\qquad +\left( \cos(|\vec{k}|\Delta t)+
|\vec{k}|\Delta t \sin(|\vec{k}|\Delta
t)\right)\left(\frac{\pi}{2}+\mathrm{si}(2|\vec{k}|\Delta
t)\right)-2\sin(|\vec{k}|\Delta t) \Bigg] \,
\end{eqnarray}
Note that the $\theta(\Delta t)$ appearing above commutes with  one of the $(\partial^{2}_{t}+|\vec{k}|^{2})$ operators because the term in square brackets is proportional to $(\Delta
t)^2$ when $\Delta t \rightarrow 0$.
Accordingly, by taking one $(\partial^{2}_{t}+|\vec{k}|^{2})$, we have
\begin{equation}\label{RetardedSelfMassFinalResult}
\imath M_{\phi,\mathrm{ren}}^{\mathrm{r}}(|\vec{k}|, t, t') =
\frac{g^{2}}{8 \pi^{2} |\vec{k}| } (\partial^{2}_{t}+|\vec{k}|^{2})^2\theta(\Delta
t) \Bigg[  \cos(|\vec{k}| \Delta t)
\left(\frac{\pi}{2}+\mathrm{si}(2|\vec{k}|\Delta t)\right) - \sin(|\vec{k}| \Delta
t) \left(\mathrm{ci}(2|\vec{k}|\Delta t) - \gamma_{E} -
\ln\left(\frac{|\vec{k}|}{2\mu^{2}\Delta t}\right)\! \right)\!\Bigg]\!
\end{equation}
We next take the Fourier transform of the above equation with respect to time as well
\begin{eqnarray}
\imath M^{\mathrm{r}}_{\phi,\mathrm{ren}}(k) &=&
\int_{-\infty}^{\infty}\mathrm{d}\Delta t e^{\imath k^{0}\Delta t}
\imath M^{\mathrm{r}}_{\phi,\mathrm{ren}}(|\vec{k}|,t,t^\prime) \label{Fourierretarded1} \nonumber \\
&=& - \frac{g^{2}}{16\pi^{2}|\vec{k}|} (-(k^{0})^{2}+|\vec{k}|^{2})^2
\int_{0}^{\infty}\mathrm{d}\Delta t \Bigg[
e^{\imath(k^{0}+|\vec{k}|)\Delta t}
\left\{-\imath\left(\mathrm{ci}(2|\vec{k}|\Delta t)-\ln(2|\vec{k}|\Delta
t)-\gamma_{E}\right) -\frac{\pi}{2}-\mathrm{si}(2|\vec{k}|\Delta t) \right\} \nonumber \\
&& \qquad\qquad\qquad\qquad\qquad\qquad\,\,\, +
e^{\imath(k^{0}-|\vec{k}|)\Delta t}
\left\{\imath\left(\mathrm{ci}(2|\vec{k}|\Delta t)-\ln(2|\vec{k}|\Delta
t)-\gamma_{E}\right) -\frac{\pi}{2}-\mathrm{si}(2|\vec{k}| \Delta
t) \right\} \nonumber\\
&&\qquad\qquad\qquad\qquad \qquad\qquad\qquad\qquad\qquad\quad -2
\imath \ln\left(2\mu\Delta t
\right)\left(e^{\imath(k^{0}+|\vec{k}|+\imath\epsilon)\Delta t} -
e^{\imath(k^{0}-|\vec{k}|+\imath\epsilon)\Delta t}\right) \Bigg] 
\end{eqnarray}
where we have introduced $\epsilon=0^+$  at necessary places, in order to regularise the integral. In order to evaluate the above equation further, we make use of the following equations \cite{AS}
\begin{subequations}
\begin{equation}
\int_{0}^{\infty}\mathrm{d}z \ln(\beta z) e^{\imath \alpha z} =
-\frac{\imath}{\alpha}\left[\ln\left(\frac{-\imath \alpha +
\epsilon}{\beta}\right)+\gamma_{E}\right]
 \label{Fourierretarded2} 
\end{equation}
\begin{equation}
e^{\imath(k^{0} \pm |\vec{k}|)\Delta t} = \frac{-\imath}{k^{0} \pm
|\vec{k}|}\partial_{t} e^{\imath(k^{0} \pm |\vec{k}|)\Delta t}
\label{Fourierretarded3} 
\end{equation}
\end{subequations}
Using these, \ref{Fourierretarded1} evaluates to
\begin{eqnarray}
\imath M^{\mathrm{r}}_{\phi,\mathrm{ren}}(k)&=&-
\frac{g^{2}}{16|\vec{k}|\pi^{2}}(-k_{0}^{2}+|\vec{k}|^{2}) \Bigg[ 2(k^{0}-|\vec{k}|)\left(
\ln\left(\frac{-\imath(k^{0}+|\vec{k}|)+\epsilon}{2\mu}\right) +
\gamma_{E}\right) \nonumber \\ && - 2(k^{0}+|\vec{k}|)\left(
\ln\left(\frac{-\imath(k^{0}-|\vec{k}|)+\epsilon}{2\mu}\right) +
\gamma_{E}\right) 
 + \int_{0}^{\infty}\mathrm{d}\Delta t \frac{2
k^{0}}{\Delta t}\left(e^{\imath(k^{0}+|\vec{k}|)\Delta t} -
e^{\imath(k^{0}-|\vec{k}|)\Delta t}\right)\Bigg] 
\label{Fourierretarded4}
\end{eqnarray}
To evaluate the remaining integrals, we use, for  real $\alpha , \beta $, \cite{AS}
\begin{equation}
\int_{0^+}^{\infty}\mathrm{d}\Delta t\left[
\frac{\cos(\alpha\Delta t)-1}{\Delta t} - \frac{\cos(\beta\Delta
t)-1}{\Delta t}\right] = \ln\left\vert\frac{\beta}{\alpha}\right\vert
\label{Fourierretarded5} 
\end{equation}
Putting these all in together, we obtain
\begin{equation}
\imath M^{\mathrm{r}}_{\phi,\mathrm{ren}}(k) =
\frac{g^{2}}{8
\pi^{2}}(-k_{0}^{2}+|\vec{k}|^{2})\left[\ln\left(\frac{-k_{0}^{2}+|\vec{k}|^{2}-\imath
\mathrm{sgn}(k^{0})\epsilon}{4\mu^{2}}\right)+2\gamma_{E}\right]
\label{Fourierretarded6}
\end{equation}
The additive constant  $2\gamma_E$ appearing in the above equation  can be further  absorbed in a scalar field strength renormalisation counterterm as of \ref{A}.
Thus we have the final expression
\begin{equation}
\imath M^{\mathrm{r}}_{\phi,\mathrm{ren}}(k) =
\frac{g^{2} k^2}{8
\pi^{2}}\ln\left(\frac{k^2-\imath
\mathrm{sgn}(k^{0})\epsilon}{4\mu^{2}}\right)
\label{Fourierretarded6}
\end{equation}
%
We next wish to identify the self-energies $\imath
M^{+-}_{\phi}(k)$ and $\imath M^{-+}_{\phi}(k)$, corresponding to the Wightman functions. The simplest way to achieve this is to use the retarded self-energy \ref{Fourierretarded6}, and $\imath
M^{++}_{\phi,\mathrm{ren}}(k)$  derived
in \ref{A}. The relationship between $\imath
M^{++}_{\phi,\mathrm{ren}}(k)$, $\imath
M^{--}_{\phi,\mathrm{ren}}(k)$, $\imath M^{\mathrm{r}}_{\phi,\mathrm{ren}}(k)$, $\imath
M^{+-}_{\phi}(k)$ and $\imath
M^{-+}_{\phi}(k)$ is given by the renormalised version of the equation appearing below \ref{FourierWightmanb1},  
\begin{equation}\label{kk}
\imath M^{\mathrm{r}}_{\phi,\mathrm{ren}}(k) = \imath
M^{++}_{\phi,\mathrm{ren}}(k) - \imath
M^{+-}_{\phi}(k)=\imath
M^{-+}_{\phi}(k) - \imath M^{--}_{\phi,\mathrm{ren}}(k)
\end{equation} 
 Using now \ref{renM++}, \ref{renM--}, \ref{Fourierretarded6} and \ref{kk}, we  find out the Wightman self-energies
\begin{subequations}
\label{FourierWightman2}
\begin{eqnarray}
\imath M^{+-}_{\phi}(k) &=& - \frac{\imath g^{2}k^2}{4\pi}\theta(-k^{0}-|\vec{k}|) \label{FourierWightman2a} \\
\imath M^{-+}_{\phi}(k) &=& - \frac{\imath
g^{2}k^{2}}{4\pi}\theta(k^{0}-|\vec{k}|) \label{FourierWightman2b} 
\end{eqnarray}
\end{subequations}

Likewise, we can find out the the renormalised advanced self-energy by using the renormalised version of the equation appearing below  \ref{FourierAdvanced1},
\begin{equation}
    \imath M^{\mathrm{a}}_{\phi,\mathrm{ren}}(k)= \imath
M^{++}_{\phi,\mathrm{ren}}(k) - \imath
M^{-+}_{\phi}(k) =  \imath M^{+-}_{\phi}(k) - \imath
M^{--}_{\phi,\mathrm{ren}}(k)
\label{advanced1}
\end{equation}
Using \ref{renM++}, \ref{renM--} and \ref{FourierWightman2}, we have
\begin{equation}
    \label{Fourieradvance6}
    \imath M^{\mathrm{a}}_{\phi,\mathrm{ren}}(k) =
\frac{g^{2}k^2}{8
\pi^{2}}\ln\left(\frac{k^2+\imath
\mathrm{sgn}(k^{0})\epsilon}{4\mu^{2}}\right)
\end{equation}
Being equipped with all these, we have computed the renormalised statistical propagator in  momentum space as quoted in the main text,~\ref{statprop1'}.\\

\noindent
 Finally, as a check of consistency, from \ref{FourierAdvanced1},  \ref{FourierWightman1}, \ref{Fourierretarded6} and \ref{Fourieradvance6}, we compute the Wightman functions,
\begin{subequations}
\label{FourierWightman3}
\begin{eqnarray}
\imath \Delta^{+-}_{\phi}(k) &=& \imath \theta(-k^{0}-|\vec{k}|)
\Bigg[\frac{1}{k^2+m^{2}+\frac{g^{2}k^2}{8
\pi^{2}}\ln\left(\frac{k^2}{4\mu^{2}}
\right)- \frac{\imath
g^{2}k^2}{8\pi}\mathrm{sgn}(k^{0})\theta(k_{0}^{2}-|\vec{k}|^{2})} \nonumber\\
&& \qquad\qquad\qquad -
\frac{1}{k^2+m^{2}+\frac{g^{2}k^2}{8
\pi^{2}}\ln\left(\frac{k^2}{4\mu^{2}}\right)
+ \frac{\imath
g^{2}k^2}{8\pi}\mathrm{sgn}(k^{0})\theta(k_{0}^{2}-|\vec{k}|^{2})} \Bigg]  \\ \label{FourierWightman3a} 
\imath \Delta^{-+}_{\phi}(k) &=& - \imath \theta(k^{0}-|\vec{k}|)
\Bigg[\frac{1}{k^2+m^{2}+\frac{g^{2}k^2}{8
\pi^{2}}\ln\left(\frac{k^2}{4\mu^{2}}\right)- \frac{\imath
g^{2}k^2}{8\pi}\mathrm{sgn}(k^{0})\theta(k_{0}^{2}-|\vec{k}|^{2})} \nonumber \\
&& \qquad\qquad\qquad -
\frac{1}{k^2+m^{2}+\frac{g^{2}k^2}{8
\pi^{2}}\ln\left(\frac{k^2}{4\mu^{2}}
\right)+ \frac{\imath
g^{2}k^2}{8\pi}\mathrm{sgn}(k^{0})\theta(k_{0}^{2}-|\vec{k}|^{2})} \Bigg] 
\label{FourierWightman3b} 
\end{eqnarray}
\end{subequations}
Using 
\begin{equation}
    \lim_{\epsilon \to 0} \frac{1}{x\pm i \epsilon}= \text{PV} \frac{1}{x} \mp i \pi \delta(x), 
\end{equation}
it is easy to see that in the limit $g \rightarrow 0$, the  above expressions agree
with the free Wightman functions
 \begin{subequations}
 \label{VacuumPropagator}
 \begin{eqnarray}
 \imath\Delta_{\phi}^{+-}(k) &=& 2\pi
 \delta(k^2+m^2) \theta(-k^{0})
 \label{VacuumPropagator+-}
 \\
 \imath\Delta_{\phi}^{-+}(k) &=& 2\pi
 \delta(k^2+m^2) \theta(k^{0}) 
 \label{VacuumPropagator-+}
 \end{eqnarray}
 \end{subequations}

\end{document}